\documentclass[sigconf, screen]{acmart}

\usepackage[linesnumbered,ruled]{algorithm2e}
\usepackage{algorithmicx}
\usepackage{algpseudocode}    
\usepackage{flushend}
\usepackage{amsmath}
\usepackage{epsfig,color}
\usepackage{graphicx}
\usepackage[utf8]{inputenc}   
\usepackage[english]{babel} 
\inputencoding{latin1}
\usepackage{mdwlist}
\usepackage{paralist} 
\usepackage{listings}
\usepackage{multirow}
\usepackage{microtype}
\usepackage{subcaption}
\usepackage{subfloat}
\usepackage{xcolor}
\definecolor{darkblue}{rgb}{0.0, 0.0, 0.55}

\usepackage{tikz}
\usetikzlibrary{positioning}
\usetikzlibrary{calc}
\usetikzlibrary{matrix}

\usepackage{wrapfig}

\usepackage{pgfplots}
\usepackage{pgfplotstable}
\pgfplotsset{compat = 1.3}
\pgfplotsset{
	legend image code/.code={
		\draw[mark repeat=2,mark phase=2]
		plot coordinates {
			(0cm,0cm)
			(0.15cm,0cm)        
			(0.3cm,0cm)         
		};%
	}
}

\usepackage{amsthm}
\theoremstyle{definition}

\let\oldnl\nl
\newcommand{\nonl}{\renewcommand{\nl}{\let\nl\oldnl}}

\newcommand{\ignore}[1]{}

\definecolor{mygreen}{rgb}{0,0.4,0.13}
\definecolor{mygray}{rgb}{0.5,0.5,0.5}
\definecolor{mymauve}{rgb}{0.58,0,0.82}
\definecolor{dkgreen}{rgb}{0,0.4,0.13}

\algnewcommand\algorithmicforeach{\textbf{for each}}
\algdef{S}[FOR]{ForEach}[1]{\algorithmicforeach\ #1\ \algorithmicdo}

\newcommand\Name{\textsf{SC-Eliminator}}

\setcopyright{acmcopyright}
\acmPrice{15.00}
\acmDOI{10.1145/3213846.3213851}
\acmYear{2018}
\copyrightyear{2018}
\acmISBN{978-1-4503-5699-2/18/07}
\acmConference[ISSTA'18]{27th ACM SIGSOFT International Symposium  on  Software Testing and Analysis}{July 16--21, 2018}{Amsterdam, Netherlands}

\begin{document}

\title{Eliminating Timing Side-Channel Leaks using Program Repair}

\author{Meng Wu}
\affiliation{%
  \institution{Virginia Tech}
  \city{Blacksburg}
  \state{VA}
  \country{USA}
}

\author{Shengjian Guo}
\affiliation{%
  \institution{Virginia Tech}
  \city{Blacksburg}
  \state{VA}
  \country{USA}
}

\author{Patrick Schaumont}
\affiliation{%
  \institution{Virginia Tech}
  \city{Blacksburg}
  \state{VA}
  \country{USA}
}

\author{Chao Wang}
\affiliation{%
  \institution{University of Southern California}
  \city{Los Angeles}
  \state{CA}
  \country{USA}
}

\begin{abstract}

We propose a method, based on program analysis and transformation, for
eliminating timing side channels in software code that implements
security-critical applications.  Our method takes as input the
original program together with a list of secret variables (e.g.,
cryptographic keys, security tokens, or passwords) and returns the
transformed program as output.  The transformed program is guaranteed
to be functionally equivalent to the original program and free of
both \emph{instruction}- and \emph{cache}-timing side channels.
Specifically, we ensure that the number of CPU cycles taken to execute
any path is independent of the secret data, and the cache behavior of
memory accesses, in terms of hits and misses, is independent of the
secret data.
We have implemented our method in LLVM and validated its effectiveness
on a large set of applications, which are cryptographic libraries with
19,708 lines of C/C++ code in total.  Our experiments show the method
is both scalable for real applications and effective in eliminating
timing side channels.

\end{abstract}



\maketitle

\section{Introduction}

Side-channel attacks have become increasingly relevant to a wide range
of applications in distributed systems, cloud computing and the
Internet of things (IoT) where timing characteristics may be exploited
by an adversary to deduce information about secret data, including
cryptographic keys, security tokens and
passwords~\cite{Kocher96,KocherJJ99,BrumleyB05,Mowery2012,liu2016,vattikonda2011}.
%
%
%
%
Generally speaking, timing side channels exist whenever the time taken
to execute a piece of software code depends on the values of secret
variables.  In this work, we are concerned with two types of timing
side-channels: instruction-related and cache-related.
By \emph{instruction}-related timing side channels, we mean the number
or type of instructions executed along a path may differ depending on
the values of secret variables, leading to differences in the number
of CPU cycles.
By \emph{cache}-related timing side channels, we mean the memory
subsystem may behave differently depending on the values of secret
variables, e.g., a cache hit takes few CPU cycles but a miss takes
hundreds of cycles.

Manually analyzing the timing characteristics of software code is
difficult because it requires knowledge of not only the application
itself but also the micro-architecture of the computer, including the
cache configuration and how software code is compiled to machine code.
Even if a programmer is able to conduct the aforementioned analysis
manually, it would be too labor-intensive and error-prone in practice:
with every code change, the software has to be re-analyzed and
countermeasure has to be re-applied to ensure a uniform execution time
for all possible values of the secret variables.
It is also worth noting that straightforward countermeasures such as
noise injection (i.e., adding random delay to the execution) do not
work well in practice, because noise can be removed using
well-established statistical analysis
techniques~\cite{Kocher96,KocherJJ99}.

Thus, we propose an fully automated method for mitigating timing side
channels.  Our method relies on static analysis to identify, for a
program and a list of \emph{secret} inputs, the set of variables whose
values depend on the secret inputs.  To decide if these
\emph{sensitive} program variables lead to timing leaks, we check if
they affect unbalanced conditional jumps (instruction-related timing
leaks) or accesses of memory blocks spanning across multiple cache
lines (cache-related timing leaks).
Based on results of this analysis, we perform code transformations to
mitigate the leaks, by equalizing the execution time.
Although our framework is general enough for a broad range of
applications, in this work, we focus on implementing a software tool
based on LLVM~\cite{LLVM} and evaluating its effectiveness on real
cryptographic software.

\begin{figure}
\centering

\vspace{-1ex}
\includegraphics[scale=.65]{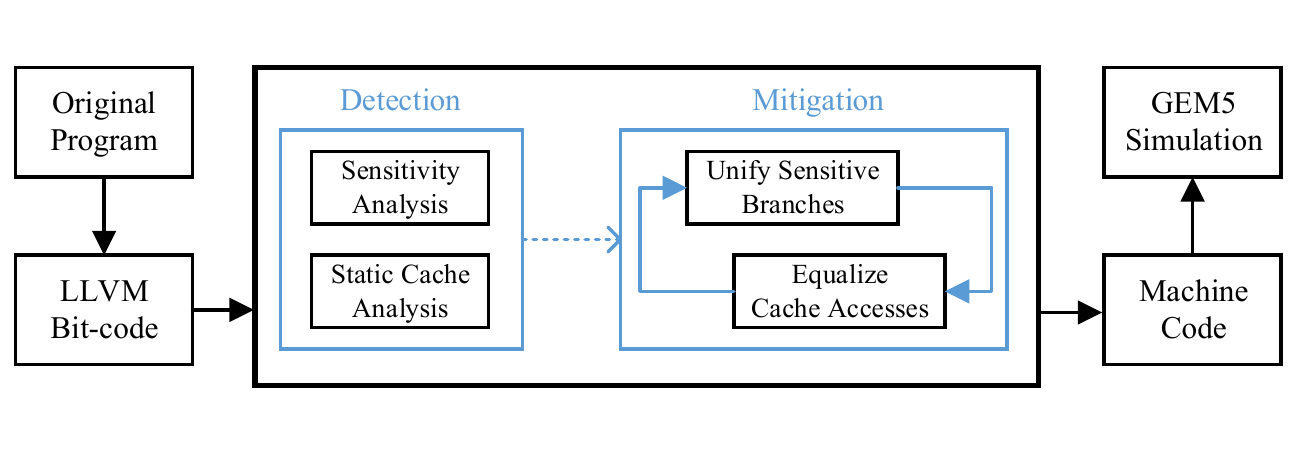}

\vspace{-4ex}
\caption{\Name{}: a tool for detecting and mitigating both \emph{instruction}- and \emph{cache}-timing side channels.}
\label{fig:diagram}
\vspace{-2ex}
\end{figure}

Figure~\ref{fig:diagram} shows the overall flow of our tool, \Name{},
whose input consists of the program and a list of secret variables.
First, we parse the program to construct its intermediate
representation inside the LLVM compiler.  Then, we conduct a series of
static analyses to identify the \emph{sensitive} variables and timing
leaks associated with these variables.  Next, we conduct two types of
code transformations to remove the leaks.  One transformation aims to
eliminate the differences in the execution time caused by unbalanced
conditional jumps, while the other transformation aims to eliminate
the differences in the number of cache hits/misses during the accesses
of look-up tables such as S-Boxes.

Conceptually, these transformations are straightforward: If we
equalize the execution time of both sensitive conditional statements
and sensitive memory accesses, there will be no instruction- or
cache-timing leaks.  However, since both transformations adversely
affect the runtime performance, they must be applied judiciously to
remain practical.  Thus, a main technical challenge is to develop
analysis techniques to decide \emph{when} these countermeasures are
\emph{not needed} and thus can be skipped safely.

Toward this end, we propose a \emph{static sensitivity analysis} to
propagate \emph{sensitivity} tags from user-annotated (secret) inputs
to other parts of the program.  The goal is to identify all variables
that \emph{may} depend transitively on the secret inputs.  Since the
analysis is static and thus has to be conservative, it
detects \emph{potential} timing leaks, e.g., unbalanced branches
guarded by sensitive variables.
We also propose a \emph{static cache analysis} to identify the set of
program locations where memory accesses always lead to cache hits.
This \emph{must-hit} analysis~\cite{FerdinandW98,FerdinandW99},
following the general framework of abstract
interpretation~\cite{CousotC77}, is designed to be conservative in
that a reported must-hit is guaranteed to be a hit along all
paths. Thus, it can be used by our tool to skip redundant mitigations.

To demonstrate that timing leaks reported by our tool are real and to
evaluate the accuracy of our static analyses, we also compile
the original and mitigated software to machine code and carefully
analyze their timing characteristics using GEM5~\cite{GEM5}, a
cycle-accurate micro-architectural CPU simulator.  Specifically, given
two values of a secret variable, denoted $k_1$ and $k_2$, we first run
the original program $P$ to show that the number of CPU cycles indeed
varies depending on the secret data; that is, $\exists
k_1,k_2: \tau(P,k_1) \neq \tau(P,k_2)$, where $\tau()$ denotes the
execution time. We then run the mitigated program $P'$ to show that,
after our mitigation, the execution time has been equalized along all
paths and for all inputs; that is, $\forall k_1,k_2: \tau(P',k_1)
= \tau(P',k_2)$.

Our method differs from recent
techniques~\cite{ChenFD17,SousaD16,AntonopoulosGHK17} for detecting
timing leaks or proving their absence: these techniques focus only
on instruction-related timing leaks but ignore the cache.  There are 
techniques that consider cache side
channels~\cite{KopfMO12,DoychevFKMR13,DoychevKMR15,ChuJM16,BasuC17,TouzeauMMR17,Chattopadhyay17,WangWLZW17}
but they focus only on leak detection as opposed to mitigation.
Our mitigation method is fundamentally different from techniques that
mitigate timing leaks~\cite{nagami2008,bortz2007,schinzel2011,Hu91} by
hiding them, e.g., by adding random delays; such countermeasures can
be easily defeated using well-known statistical analysis
techniques~\cite{Kocher96,KocherJJ99}.
Finally, since our method is a software-only solution, it is more
flexible and more widely applicable than techniques that require
hardware support (e.g.,~\cite{ZhangAM12} and \cite{AwekeA17}).

We have implemented our method in a software tool and evaluated it on
many cryptographic libraries, including \emph{Chronos}~\cite{Chronos},
a real-time Linux kernel;
\emph{FELICS}~\cite{DinuBGKCP2015}, a lightweight cryptographic 
systems for IoT devices; \emph{SUPERCOP}~\cite{Supercop}, a toolkit
for measuring the performance of cryptographic algorithms;
\emph{Botan}~\cite{Botan}, a cryptographic library 
written in C++11; and \emph{Libgcrypt}~\cite{Libgcrypt}, the GNU
library.  In total, they have 19,708 lines of C/C++ code.
Our experiments show the tool is scalable for these real applications:
in all cases, the static analysis took only a few seconds while the
transformation took less than a minute.  Furthermore, the mitigated
software have only moderate increases in code size and runtime
overhead.  Finally, with GEM5 simulation, we were able to confirm that
both instruction- and cache-timing leaks were indeed eliminated.

To summarize, this paper makes the following contributions:
\begin{itemize}
\item
We propose a static analysis and transformation based method for
eliminating instruction- and cache-timing side channels.
\item
We implement the proposed method in a software tool based on LLVM,
targeting cryptographic software written in C/C++.
\item
We evaluate our tool on a large number of applications to demonstrate
its scalability and effectiveness.
\end{itemize}

The remainder of this paper is organized as follows. First, we use
examples to illustrate instruction- and cache-timing side channels in
Section~\ref{sec:motivation}, before defining the notations in
Section~\ref{sec:prelim}.  We present our methods for detecting timing
leaks in Section~\ref{sec:detection} and for mitigating timing leaks
in Sections~\ref{sec:mitigation1} and ~\ref{sec:mitigation2}.  We
present our experimental results in Section~\ref{sec:experiment},
review the related work in Section~\ref{sec:related}, and finally,
give our conclusions in Section~\ref{sec:conclusion}.

\section{Motivation}
\label{sec:motivation}

In this section, we use real examples to illustrate various types of timing
leaks in cryptographic software.

\subsection{Conditional Jumps Affected by Secret Data}

An unbalanced if-else statement whose condition is affected by secret
data may have side-channel leaks, because the \emph{then}-
and \emph{else}-branches will have different execution time.
Figure~\ref{fig:branchleak} shows the C code of a textbook
implementation of a \texttt{3-way} cipher~\cite{AppliedCryp}, where
the variable $a$ is marked as secret and it affects the execution time
of the if-statements.  By observing the timing variation, an adversary
may be able to gain information about the bits of $a$.

To remove the dependencies between execution time and secret data, one
widely-used approach is equalizing the branches by cross-copying~\cite{Agat00,MolnarPSW05,KopfM07} as
illustrated by the code snippet in the middle of
Figure~\ref{fig:branchleak}: the auxiliary variable
\texttt{dummy\_b[3]} and some assignments are added to make both branches 
contain the same number and type of instructions.
Unfortunately, this approach does not always work in practice, due to
the presence of hidden states at the micro-architectural levels and
related performance optimizations inside modern CPUs (e.g.,
instruction caching and speculative execution) --
we have confirmed this limitation by analyzing the mitigated code
using GEM5, the details of which are described as follows.

\begin{figure}
\vspace{1ex}
\centering
\begin{minipage}{0.960\linewidth}

\begin{lstlisting}
void mu(int32_t *a) {            // original version
	int i;
	int32_t b[3];
	b[0] = b[1] = b[2] = 0;
	for (i=0; i<32; i++) {
		b[0] <<= 1; b[1] <<= 1; b[2] <<= 1;
		if(a[0]&1) b[2] |= 1;     // leak
		if(a[1]&1) b[1] |= 1;     // leak
		if(a[2]&1) b[0] |= 1;     // leak
		a[0] >>= 1; a[1] >>= 1; a[2] >>= 1;
	}
	a[0] = b[0]; a[1] = b[1]; a[2] = b[2];
}\end{lstlisting}

\begin{lstlisting}
	// mitigation #1: equalizing the branches
	int32_t dummy_b[3];
	dummy_b[0] = dummy_b[1] = dummy_b[2] = 0;
	... 
		dummy_b[0] <<= 1; dummy_b[1] <<= 1; dummy_b[2] <<= 1;
		... 
		if(a[0]&1) b[2]|=1; else dummy_b[2]|=1;   
		if(a[1]&1) b[1]|=1; else dummy_b[1]|=1;   
		if(a[2]&1) b[0]|=1; else dummy_b[0]|=1;   
\end{lstlisting}

\begin{lstlisting}
	// mitigation #2: removing the branches
    b[2] = CTSEL(a[0]&1, b[2]|1, b[2]);
    b[1] = CTSEL(a[1]&1, b[1]|1, b[1]);
    b[0] = CTSEL(a[2]&1, b[0]|1, b[0]);
\end{lstlisting}

\end{minipage}

\vspace{-1ex}
\caption{Example code from a real cipher with timing leakage, together with two different mitigation approaches.}
\label{fig:branchleak}
\vspace{-1ex}
\end{figure}

We compiled the mitigated program shown in the middle of
Figure~\ref{fig:branchleak} and, by carefully inspecting the machine
code, made sure that all conditional branches indeed had the same
number (and type) of instructions.  Then, we ran the top-level program
on GEM5 with two different cryptographic keys: $k_1$ has 1's in all 96
bits whereas $k_2$ has 0's in all 96 bits. Our GEM5 simulation results
showed significant timing differences: 88,014 CPU cycles for $k_1$
versus 87,624 CPU cycles for $k_2$.  Such timing variation would
allow attackers to gain information about the secret key.

Therefore, in the remainder of this paper, we avoid the aforementioned
approach while focusing on a safer alternative:
replacing \emph{sensitive branches} with functionally-equivalent,
constant-time, and \emph{branch-less} assignments shown at the
bottom of Figure~\ref{fig:branchleak}.
Specifically, \emph{CTSEL(c,t,e)} is an LLVM intrinsic we added to
ensure the selection of either $t$ or $e$, depending on the predicate
$c$, is done in constant time.  For different CPU architectures, this
intrinsic function will be compiled to different machine codes to
obtain the best performance possible (see
Section~\ref{sec:mitigation1} for details).
Because of this, our mitigation adds little runtime overhead: the
mitigated program requires only 90,844 CPU cycles for both $k_1$ and
$k_2$.


Note that we cannot simply rely on C-style conditional
assignment \texttt{r=(c?t:e)} or the LLVM \emph{select} instruction
because neither guarantees constant-time execution.  Indeed, LLVM may
transform both to conditional jumps, e.g., when $r$ is
of \texttt{char} type, which may have the same residual timing leaks
as before.  In contrast, our use of the new \emph{CTSEL} intrinsic
avoids the problem.

\subsection{Table Lookups Affected by Secret Data}

When an index used to access a lookup table (LUT) depends on the
secret data, the access time may vary due to the behavior of cache
associated with the memory block. Such cache-timing leaks have been
exploited in block
ciphers~\cite{Spreitzer2013,Mowery2012,Gullasch2011} that, for
efficiency reasons, implement \texttt{S-Boxes} using lookup tables.
Figure~\ref{fig:TLU} shows the \texttt{subBytes} function of the AES
cipher in FELICS~\cite{DinuBGKCP2015}, which substitutes each byte of
the input array (\texttt{block}) with the precomputed byte stored
in \texttt{sbox}.  Thus, the content of \texttt{block}, which depends
on secret data, may affect the execution time.
For example, when all sixteen bytes of \texttt{block}
are \texttt{0x0}, meaning \texttt{sbox[0]} is always accessed, there
will be one cache miss followed by fifteen hits; but when all sixteen
bytes of \texttt{block} differ from each other, there may be 256/64 =
4 cache misses (if we assume 64 bytes per cache line).

\begin{figure}
\vspace{1ex}
\centering
\begin{minipage}{0.960\linewidth}
\begin{lstlisting}
const uint8_t sbox[256] = { 0x63, 0x7c, 0x77, 0x7b, 0xf2, 0x6b, 0x6f, 0xc5, 
	0x30, 0x01, 0x67, 0x2b, 0xfe, 0xd7, 0xab, 0x76,	...};
void subBytes(uint8_t *block) {
	uint8_t i;
	for (i = 0; i < 16; ++i) {
		block[i] = sbox[block[i]];
	}
}
\end{lstlisting}
\end{minipage}

\vspace{-2ex}
\caption{Example for accessing the lookup table.}
\label{fig:TLU}
\vspace{-1ex}
\end{figure}

\begin{figure}
\centering
\begin{minipage}{0.960\linewidth}
\begin{lstlisting}
		//mitigation #3: replacing block[i] = sbox[block[i]];
		block_i = block[i];
		for (j=0; j < 256; j++) {
			sbox_j = sbox[j];
			val = (block_i == j)? sbox_j : block_i;
		}
		block[i] = val;
\end{lstlisting}
\end{minipage}

\vspace{-2ex}
\caption{Countermeasure: reading all the elements.}
\label{fig:TLUfix}
\vspace{-1ex}
\end{figure}

Mitigating cache-timing leaks differs from mitigating
instruction-timing leaks.  Generally speaking, the level of
granularity depends on the threat model (i.e., what the attacker can
and cannot do).
For example, if we add, as camouflage, accesses of all elements
of \texttt{sbox[256]} to each original read of \texttt{sbox[]}, as
shown in Figure~\ref{fig:TLU}, it would be impossible for attackers to
guess which is the desired element.  Since each original loop
iteration now triggers the same number of LUT accesses, there is no
longer timing variation.

\begin{figure}
\centering
\begin{minipage}{0.960\linewidth}
\begin{lstlisting}
		//mitigation #4: replacing block[i] = sbox[block[i]];
		block_i = block[i];
		for (j=block_i % CLS; j < 256; j+=CLS) {
			sbox_j = sbox[j];
			val = (block_i == j)? sbox_j : block_i;
		}
		block[i] = val;
\end{lstlisting}
\end{minipage}

\vspace{-2ex}
\caption{Countermeasure: reading all cache lines.}
\label{fig:TLUfix2}
\vspace{-1ex}
\end{figure}

However, the high runtime overhead may be unnecessary, e.g., when
attackers cannot observe the timing variation of each loop iteration.
If, instead, the attackers can only observe differences in the cache
line associated with each write to \texttt{block[i]}, it suffices to
use the approach in Figure~\ref{fig:TLUfix2}.  Here, \texttt{CLS}
denotes the cache line size (64 bytes in most modern CPUs).
Note there is a subtle difference between this approach and the naive
preloading (Figure~\ref{fig:TLUfix3}): the latter would be vulnerable
to \texttt{Flush+Reload} attacks~\cite{OsvikST06,YaromF14}.
For example, the attackers can carefully arrange the \texttt{Flush}
after \texttt{Preload} is done, and then perform \texttt{Reload} at
the end of the victim's computation; this is possible
because \texttt{Preload} triggers frequent memory accesses that are
easily identifiable by an attacker.
In contrast, the approach illustrated in Figure~\ref{fig:TLUfix2} can
avoid such attacks.

\begin{figure}
\centering
\begin{minipage}{0.960\linewidth}
\begin{lstlisting}
	//mitigation #5: preloading sbox[256]
	for (j =0; j < 256; j+=CLS)
		temp = sbox[j];
	//access to sbox[...] is always a hit
	for (i = 0; i < 16; ++i) {
		block[i] = sbox[block[i]];
	}
\end{lstlisting}
\end{minipage}

\vspace{-2ex}
\caption{Countermeasure: preloading all cache lines.}
\label{fig:TLUfix3}
\vspace{-1ex}
\end{figure}

If the attackers can only measure the total execution time of a
program, our mitigation can be more efficient than
Figures~\ref{fig:TLUfix3} and \ref{fig:TLUfix2}: For example, if the
cache is large enough to hold all elements, preloading would
incur \texttt{256/CLS}=4 cache misses, but all subsequent accesses
would be hits.  This approach will be illustrated in
Figure~\ref{fig:TLUfix4} (Section~\ref{sec:mitigation2}). 
However, to safely apply such optimizations, we need to make sure the
table elements never get evicted from the cache. For simple loops,
this would be easy.  But in real applications, loops may be complex,
e.g., containing branches, other loops, and function calls, which
means in general, a sound static program analysis procedure is needed
(see Section~\ref{sec:optimization}) to determine whether a lookup
table access is a MUST-HIT.


\subsection{Idiosyncratic Code Affected by Secret Data}
\label{OpExpansion}

For various reasons, certain operations in cryptographic software are
often implemented using a series of simpler but
functionally-equivalent operations.
For example, the shift operation \texttt{(X<<C)} may be implemented
using a sensitive data-dependent loop with
additions: \texttt{for(i=0;i<C;i++) \{X += X;\}} because some targets
(e.g. MSP430) do not support multi-bit shifts.

\begin{figure}
\vspace{1ex}
\centering
\begin{minipage}{0.960\linewidth}

\begin{lstlisting}
typedef struct {
	uint32_t *xk; // the round keys
	int nr;       // the number of rounds
} rc5_ctx;
#define ROTL32(X,C) (((X)<<(C))|((X)>>(32-(C))))
void rc5_encrypt(rc5_ctx *c, uint32_t *data, int blocks) {
	uint32_t *d,*sk;
	int h,i,rc;
	d = data;
	sk = (c->xk)+2;
	for (h=0; h<blocks; h++) {
		d[0] += c->xk[0];
		d[1] += c->xk[1];
		for (i=0; i<c->nr*2; i+=2) {
			d[0] ^= d[1];
			rc = d[1] & 31;
			d[0] = ROTL32(d[0],rc);
			d[0] += sk[i];
			d[1] ^= d[0];
			rc = d[0] & 31;
			d[1] = ROTL32(d[1],rc);
			d[1] += sk[i+1];
		}
		d+=2;
		}
}
\end{lstlisting}
\end{minipage}

\vspace{-1ex}
\caption{Code snippet from RC5.c}
\label{fig:RC5}
\vspace{-2ex}
\end{figure}

One real example of such idiosyncratic code is the implementation
of \texttt{rc5\_encrypt}~\cite{AppliedCryp} shown in
Figure~\ref{fig:RC5}.  Here, the second parameter of \texttt{ROTL32()}
is aliased to the sensitive variable \texttt{c->xk}.
To eliminate the timing leaks caused by an idiosyncratic
implementation of \texttt{(X<<C)}, we must conservatively estimate the
loop bound.  If we know, for example, the maximum value of \texttt{C}
is \texttt{MAX\_C}, the data-dependent loop may be rewritten to one
with a fixed loop bound: 
\texttt{for(i=0;i<MAX\_C;++i) \{if(i<C) X += X;\}}.  
After this transformation, we can leverage the aforementioned
mitigation techniques to eliminate leaks associated with
the \texttt{if(i<C)} statement.


\section{Threat Model}
\label{sec:prelim}

We now define the threat model, as well as timing side-channel leaks
under our threat model.


We assume a \emph{less-capable} attacker who can only observe
variation of the total execution time of the victim's program with
respect to the secret data.  Since this capability is easier to obtain
than that of a \emph{more-capable} attacker, it will be more widely
applicable.  A classic example, for instance, is when the victim's
program runs on a server that can be probed and timed remotely by the
attacker using a malicious client.

We do not consider the \emph{more-capable} attacker who can directly
access the victim's computer to observe hidden states of the CPU at
the micro-architectural levels, e.g., by running malicious code to
perform Meltdown/Spectre
attacks~\cite{Lipp2018meltdown,Kocher2018spectre} or similar cache
attacks~\cite{OsvikST06,YaromF14} (Evict+Time, Prime+Probe, and
Flush+Reload).  Mitigating such attacks at the software level only
will likely be significantly more expensive --- we leave it for future
work.


Let $P$ be a program and $in = \{X,K\}$ be the input, where $X$ is
\emph{public} and $K$ is \emph{secret}.  Let $x$
and $k$ be concrete values of $X$ and $K$, respectively, and
$\tau(P,x,k)$ be the time taken to execute $P$ under $x$ and $k$.
%
%
We say $P$ is free of timing side-channel leaks if
\[ \forall x, k_1, k_2 :  \tau(P,x,k_1) = \tau(P,x,k_2) ~.\]
%
That is, the execution time of $P$ is independent of the secret input
$K$.  When $P$ has timing leaks, on the other hand, there must exist
some $x$, $k_1$ and $k_2$ such that $\tau(P,x,k_1) \neq \tau(P,x,k_2)$.

We assume $P$ is a deterministic program whose execution is fixed
completely by the input.  Let $\pi = inst_1,\ldots,inst_n$ be an
execution path, and $\tau(inst_i)$ be the time taken to execute each
instruction $inst_i$, where $1\leq i\leq n$; then, we have $\tau(\pi)
= \Sigma_{i=1}^{n} \tau(inst_i)$.

Furthermore, $\tau(inst_i)$ consists of two components:
$\tau_{cpu}(inst_i)$ and $\tau_{mem}(inst_i)$, where $\tau_{cpu}$
denotes the time taken to execute the instruction itself and
$\tau_{mem}(inst_i)$ denotes the time taken to access the memory.
For \emph{Load} and \emph{Store}, in particular, $\tau_{mem}(inst_i)$
is determined by whether the access leads to a cache hit or miss. For the
other instructions, $\tau_{mem}(inst_i) = 0$.
We want to equalize both components along all program paths -- this
will be the foundation of our leak mitigation technique.

\section{Detecting Potential Leaks}
\label{sec:detection}

Now, we present our method for detecting timing leaks, which is
implemented as a sequence of LLVM passes at the IR level.  It takes a
set of input variables marked as \textit{secret} and returns a set of
instructions whose execution may depend on these secret inputs.

\subsection{Static Sensitivity Analysis}

To identify the leaks, we need to know which program variables are
dependent of the \emph{secret} inputs --- they are
the \emph{sensitive} variables.  Since manual annotation is tedious
and error prone, we develop a procedure to perform such annotation
automatically.

\vspace{1ex}
\noindent
\textbf{Secret Source:} 
The initial set of \emph{sensitive} variables consists of the secret
inputs marked by the user.  For example, in a block cipher, the secret
input would be the cryptographic key while the plaintext would be
considered as public.

\vspace{1ex}
\noindent
\textbf{Tag Propagation:}
The \emph{sensitivity} tag is an attribute to be propagated from the
secret source to other program variables following either data- or
control-dependency transitively.
%
An example of data-dependency is the \emph{def-use} relation in
\texttt{\{b = a \& 0x80;\}} where  $b$ is marked as sensitive
because it depends on the most significant bit of $a$, the sensitive
variable.
%
An example of control-dependency is \texttt{if(a==0x10) \{b=1;\}
else \{b=0;\}} where $b$ is marked as sensitive because it depends on
whether $a$ is \texttt{0x10}.

\vspace{1ex}
\noindent
\textbf{Field-sensitive Analysis:}
To perform the static analysis defined above, we need to identify
aliased expressions, e.g., syntactically-different variables or fields
of structures that point to the same memory location.  Cryptographic
software code often has this type of pointers and structures.
For example, the ASE implementation of Chronos~\cite{Chronos} shown in
Figure~\ref{fig:fieldanalysis} demonstrates the need for
field-sensitivity during static analysis.  Here, local
pointer \texttt{key} becomes sensitive when \texttt{key[0]} is
assigned the value of another sensitive variable
\texttt{in\_key}.  Without field sensitivity, one would have to 
mark the entire structure as sensitive to avoid missing potential
leaks.  In contrast, our method performs a field-sensitive pointer
analysis~\cite{BalatsourasS16,PearceKH04} to propagate the sensitivity
tag only to relevant fields such as \texttt{key\_enc}
inside \texttt{ctx}, while avoiding fields such
as \texttt{key\_length}.  This means we can avoid marking (falsely)
the unbalanced \texttt{if (ctx->key\_length)} statement as leaky.

\begin{figure}
\vspace{1ex}
\centering
\begin{minipage}{0.960\linewidth}

	\begin{lstlisting}
struct aes_ctx {
  uint32_t key_enc[60];
  uint32_t key_length;
};
int expand_key(const uint8_t *in_key, struct aes_ctx *ctx, unsigned int key_len)
{
  uint32_t *key =ctx->key_enc;
  key[0] = *((uint32_t*)in_key);
  ...
  ctx->key_length = key_len;
  ...
  if (ctx->key_length)
    ...
}\end{lstlisting}
\end{minipage}

\vspace{-2ex}
	\caption{Example of field-sensitive pointer analysis.}
	\label{fig:fieldanalysis}
\vspace{-1ex}

\end{figure}

\ignore{ 
Although there are already plenty of existing works on pointer
analysis~\cite{BalatsourasS16}~\cite{PearceKH04} or taint
analysis~\cite{saint}, our benchmarks are not general programs but
cryptographic ciphers ranging from just 100 to 1500 lines of code
where their analysis are just too heavy. For simplicity, we implement
our just-good-enough version of sensitivity analysis as an LLVM pass
with only 300 lines of code.  
}

\subsection{Leaky Conditional Statements}

There are two requirements for a branch statement to have potential
timing leaks.  First, the condition depends on secret data. Second,
the branches are unbalanced.
Figure~\ref{fig:branchleak} shows an example, where the conditions
depend on the secret input \texttt{a} and the branches obviously are
unbalanced.
Sometimes, however, even if two conditional branches have the same
number and type of instructions, they still result in different
execution time due to hidden micro-architectural states, as we have
explained in Section~\ref{sec:motivation} and confirmed using GEM5
simulation.
Thus, to be conservative, we consider \emph{all} sensitive conditional
statements as potential leaks (regardless of whether they are 
balanced) and apply our \emph{CTSEL} based mitigation.

\subsection{Leaky Lookup-table Accesses}

The condition for a lookup-table (LUT) access to leak timing
information is that the index used in the access is sensitive.  In
practice, the index affected by secret data may cause memory accesses
to be mapped to different cache lines, some of which may have been
loaded and thus result in hits while others result in misses.
Therefore, we consider LUT accesses indexed by sensitive variables as
potential leaks, e.g., the load from \texttt{sbox} in
Figure~\ref{fig:TLU}, which is indexed by a sensitive element
of \textit{block}.

However, not all LUT accesses are leaks. For example, if the table has
already been loaded, the (sensitive) index would no longer cause
differences in the cache. This is an important optimization we perform
during mitigation --- the analysis required for deciding \emph{if
an LUT access results in a must-hit} will be presented in
Section~\ref{sec:optimization}.

\section{Mitigating Conditional Statements}
\label{sec:mitigation1}

In this section, we present our method for mitigating leaks associated
with conditional jumps. 
In contrast to existing approaches that only attempt to balance the
branches, e.g., by adding dummy
instructions~\cite{Agat00,MolnarPSW05,KopfM07}, we eliminate these branches.

\begin{algorithm}
\caption{Mitigating all sensitive conditional statements.}
\label{alg:branch}
{\footnotesize
 \SetAlgoLined
 \DontPrintSemicolon
	
                $\mathit{BranchMitigatePass}$ (Function $F$)\;
		\Begin{
			let $\mathit{DT(F)}$ be the dominator tree in the CFG of $F$;\;
			\ForEach{BasicBlock $bb \in \mathit{DT(F)}$ in DFS order}{			
				\uIf{bb is the entry of a sensitive conditional statement}{
					Standardize ($bb$);\;
					MitigateBranch ($bb$);\;	
				}				
			}
		}

}
\end{algorithm}

Algorithm~\ref{alg:branch} shows our high-level procedure implemented
as an LLVM optimization (\emph{opt}) pass: for each function $F$, we
invoke \emph{BranchMitigationPass(F)} to compute the dominator tree of
the control flow graph (CFG) associated with $F$ and then traverse the
basic blocks in a depth-first search (DFS) order.

The dominator tree is a standard data structure in compilers where
each basic block has a unique immediate dominator, and an edge from
$bb_1$ to $bb_2$ exists only if $bb_1$ is an immediate dominator of
$bb_2$.
The DFS traversal order is important because it guarantees to visit
the inner-most branches before the outer branches.  Thus,
when \emph{MitigateBranch(bb)} is invoked, we know all branches
inside \emph{bb} have been mitigated, i.e., they are either removed or
insensitive and hence need no mitigation.

Our mitigation of each conditional statement starting with $bb$
consists of two steps: (1) transforming its IR to a standardized form,
using \emph{Standardize(bb)}, to make subsequent processing easier;
and (2) eliminating the conditional jumps
using \emph{MitigateBranch(bb)}.

\subsection{Standardizing Conditional Statements}

\begin{figure}
\centering
\begin{minipage}{.3\linewidth}
\begin{lstlisting}
do {
	t += 1;
	if (x == 1)
	  break;
	x >>1;
	...
} while (x != 0);
return t;
\end{lstlisting}
\end{minipage}
\begin{minipage}{.45\linewidth}
\centering
\vspace{-3ex}
\scalebox{0.85}{
  \begin{tikzpicture}[fill=blue!20,font=\footnotesize,
  block/.style={
  	draw,
  	fill=white,
  	rectangle, 
  	minimum width={1.3cm},
  	font=\small}] 
	\node[block, align=center] (1) at (2,3) {t+=1;\\if(x==1)};
	\node[block, align=center] (2) at (1,2) {x\textgreater \textgreater1;\\while(x!=0)};
	\node[block, align=center] (3) at (2,1) {return t;};
	\draw[->] (1) -- (2);
	\draw[->] (2) -- (3);
	\draw[->] (1) -- (3);
	\draw[->] (2) to[out=-90,in=-90] (-0.25,1.75) to [out=90,in=120] (1);
	\draw[->] (2,3.7) -- (2,3.39);
  \end{tikzpicture}
 
}
\end{minipage}

\vspace{-1ex}
\caption{A not-yet-standardized conditional statement.}
\label{fig:badBranch}
\vspace{-1ex}
\end{figure}
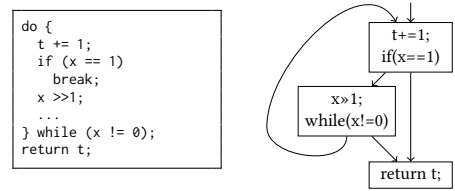

A conditional statement is standardized if it has unique entry and
exit blocks.  In practice, most conditional statements in
cryptographic software are already standardized.
However, occasionally, there may be statements that do not conform to
this requirement.  For example, in Figure~\ref{fig:badBranch}, the
conditional statement inside the while-loop is not yet standardized.
In such cases, we transform the LLVM IR to make sure it is
standardized, i.e., each conditional statement has a unique entry
block and a unique exit block.

\begin{figure}
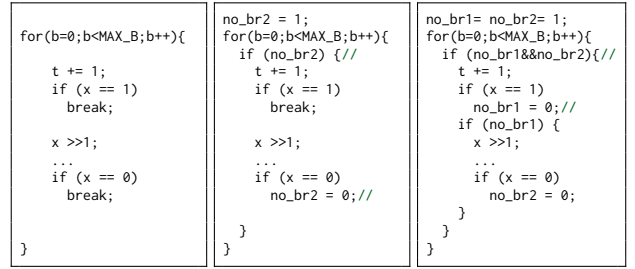

\centering
\begin{minipage}{.28\linewidth}
\begin{lstlisting}[numbers=none]

for(b=0;b<MAX_B;b++){

		t += 1;
		if (x == 1)   
		  break; 

		x >>1;
		...
		if (x == 0)
		  break;   


}
\end{lstlisting}
\end{minipage}
\hspace{.02\linewidth}
\begin{minipage}{.28\linewidth}
\begin{lstlisting}
no_br2 = 1;
for(b=0;b<MAX_B;b++){
	if (no_br2) {//
		t += 1;
		if (x == 1)   
		  break; 

		x >>1;
		...
		if (x == 0)        
		  no_br2 = 0;//

	}
} 
\end{lstlisting}
\end{minipage}
\hspace{.02\linewidth}
\begin{minipage}{.3\linewidth}
\begin{lstlisting}
no_br1= no_br2= 1; 
for(b=0;b<MAX_B;b++){
	if (no_br1&&no_br2){//
		t += 1;
		if (x == 1)   
		  no_br1 = 0;//
		if (no_br1) {
			x >>1;
			...
			if (x == 0)   
			  no_br2 = 0;
		}
	}
} 
\end{lstlisting}
\end{minipage}

\vspace{-1ex}
\caption{Standardized conditional statements (Fig.~\ref{fig:badBranch}).}
\label{fig:badBranchFix}
\vspace{-1ex}
\end{figure}

Standardization is a series of transformations as illustrated by the
examples in Figure~\ref{fig:badBranchFix}, where auxiliary variables
such as \texttt{no\_br1} and \texttt{no\_br2} are added to make the
loop bound independent of sensitive variables.  \texttt{MAX\_B} is the
bound computed by our conservative static analysis; in cryptographic
software, it is often 64, 32, 16 or 8, depending on the number of bits
of the variable \texttt{x}.

\subsection{Replacing Conditional Statements}

Given a standardized conditional statement, we perform a DFS traversal
of its dominator tree to guarantee that we always mitigate the
branches before their merge point.  The pseudo code, shown in
Algorithm~\ref{alg:branch2}, takes the entry block $bb$ as input.


\noindent
\textbf{Condition and CTSEL:}  
First, we assume the existence of \emph{CTSEL(c,t,e)}, a constant-time
intrinsic function that returns $t$ when $c$ equals \emph{true}, and
$e$ when $c$ equals \emph{false}.
Without any target-specific optimization, it may be implemented using
bit-wise operations:
\texttt{CTSEL(c,t,e) \{c$_0$=c-1; c1=$\sim$c$_0$; val= (c$_0$ \& e)|(c$_1$ \& t);\}}
--- when the variables are of 'char' type and $c$ is $true$, $c_0$
will be $0x00$ and $c_1$ will be $0xFF$; and when $c$ is $false$,
$c_0$ will be $0xFF$ and $c_1$ will be $0x00$.
With target-specific optimization, \emph{CTSEL(c,t,e)} may be
implemented more efficiently.  For example, on x86 or ARM CPUs, we may
use \emph{CMOVCC} instructions as follows:
\texttt{\{MOV val t; CMP c 0x0; CMOVEQ val e;\}} 
which requires only three instructions.
We will demonstrate through experiments (Section~\ref{sec:experiment})
that target-specific optimization reduces the runtime overhead
significantly.

\begin{algorithm}
\caption{Mitigating the conditional statement from $bb$.}
\label{alg:branch2}
{\footnotesize
 \SetAlgoLined
 \DontPrintSemicolon

		$\mathit{MitigateBranch}$ (BasicBlock $bb$)\;
		\Begin{
                Let $cond$ be the branch condition associated with $bb$;\;
		        \ForEach{Instruction $i$ in THEN branch or ELSE branch}{	
			        \uIf{$i$ is a \textbf{Store} of the value $\mathit{val}$ to the memory address $\mathit{addr}$}{
                             Let ${val'} = \textbf{\textit{CTSEL}}(cond, val, \textbf{\textit{Load}}(\mathit{addr}))$;\;  
		 	                 Replace $i$ with the new instruction $\textbf{\textit{Store}}(\mathit{val',addr})$;\;
			        }
		        }

				\ForEach{Phi Node ($\%rv \leftarrow \phi(\%rv_T,\%rv_E)$) at the merge point}{
					Let $val' = \textbf{\textit{CTSEL}}(cond,\%rv_T, \%rv_E)$;\;
					Replace the Phi Node with the new instruction ($\%rv \leftarrow val'$);\;
				}

			    Change the conditional jump to THEN branch to unconditional jump;\;  
	            Delete the conditional jump to ELSE branch;\; 
	            Redirect the outgoing edge of THEN branch to start of ELSE branch;\;

	        }
}
\end{algorithm}

\noindent
\textbf{Store Instructions:}
Next, we transform the branches.  If the instruction is a \emph{\it
Store(val,addr)} we replace it with \emph{CTSEL}.  That is,
the \emph{\it Store} instructions in \texttt{THEN} branch will only
take effect when the condition is evaluated to $true$, while
the \emph{\it Store} instructions in \texttt{ELSE} branch will only
take effect when condition is $false$.


\vspace{1ex}
\noindent
\textbf{Local Assignments:}
The above transformation is only for memory \emph{Store}, not
assignment to a register variable such as \texttt{if(cond) \{rv=val1; ...\} else \{rv=val2; ...\}} because, inside LLVM, the latter
is represented in the static single assignment (SSA) format.  Since
SSA ensures each variable is assigned only once, it is equal
to \texttt{if(cond) \{\%rv$_1$=val1; ...\} else \{\%rv$_2$=val2; ...\}} together with a \emph{Phi Node} added to the merge point of
these branches.

\vspace{1ex}
\noindent
\textbf{The Phi Nodes:}
The Phi nodes are data structures used by compilers to represent all
possible values of local (register) variables at the merge point of
CFG paths.  For $\%rv \leftarrow \phi(\%rv_T,\%rv_E)$, the variables
$\%rv_T$ and $\%rv_E$ in SSA format denote the last definitions of
$\%rv$ in the THEN and ELSE branches: depending on the condition,
$\%rv$ gets either $\%rv_T$ or $\%rv_E$.
Therefore, in our procedure, for each Phi node at the merge point, we
create an assignment from the newly created $val'$ to $\%rv$, where
$val'$ is again computed using \texttt{CTSEL}.

\begin{figure}
\vspace{1ex}
\centering
\scalebox{0.85}{  \begin{tikzpicture}[fill=blue!20,font=\footnotesize,
  block/.style={
  	draw,
  	fill=white,
  	rectangle, 
  	minimum width={1.3cm},
  	font=\small}] 
	\node[block, align=center] (Start) at (1,4.25) {Entry};
	\node[block] (If) at (0,3) {Then};
	\node[dashed, block,  align=center]  at (0,2) {blocks};
	\node[block] (ifend) at (0,1) {End\_T};
	\node[block] (Else) at (2,3) {Else};
	\node[dashed, block,  align=center]  at (2,2) {blocks};
	\node[block] (elseend) at (2,1) {End\_E};
	\node[block] (End) at (1,0) {Exit};
	\draw[->] (Start) -- node[left]{(cond)~~}    (If);
	\draw[->] (Start) -- node[right]{~~(!cond)}  (Else);
	\draw[->] (ifend) -- (End);
	\draw[->] (elseend) -- (End);
	\draw[dashed] (-0.8, 0.7) rectangle (0.8, 3.3);
	\draw[dashed] (1.2, 0.7) rectangle (2.8, 3.3);
	\node[block, align=center] (Start) at (7,4.25) {Entry};
	\node[block] (If) at (6,3) {Then};
	\node[dashed, block,  align=center]  at (6,2) {blocks};
	\node[block] (ifend) at (6,1) {End\_T};
	\node[block] (Else) at (8,3) {Else};
	\node[dashed, block,  align=center]  at (8,2) {blocks};
	\node[block] (elseend) at (8,1) {End\_E};
	\node[block] (End) at (7,0) {Exit};
	\draw[->,blue] (Start) -- node[left]{(true)~~}    (If);
	\draw[->,blue,rounded corners=10pt] (ifend) -- (6,0.25) -- (7,0.5) -- (7,3.5) -- (8,3.9) -- node[right]{~~(true)} (Else);
	\draw[->] (elseend) -- (End);
	\draw[dashed] (5.2, 0.7) rectangle (6.8, 3.3);
	\draw[dashed] (7.2, 0.7) rectangle (8.8, 3.3);
  \end{tikzpicture}
 }

\vspace{-1ex}
\caption{Removing the conditional jumps.}
\label{fig:standBranch3}
\vspace{-1ex}
\end{figure}
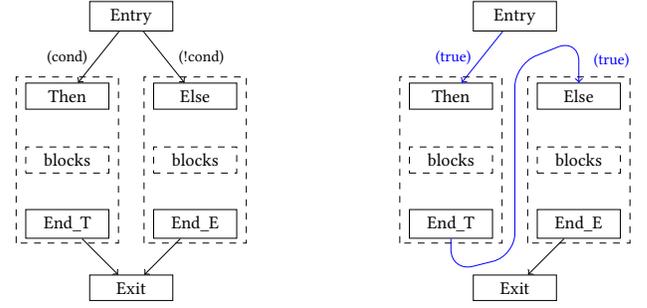

\vspace{1ex}
\noindent
\textbf{Unconditional Jumps:}
After mitigating both branches and the merge point, we can eliminate
the conditional jumps with unconditional jumps.  For the standardized
branches on the left-hand side of Figure~\ref{fig:standBranch3}, the
transformed CFG is shown on the right-hand side.

\subsection{Optimizations}

The approach presented so far still has redundancy.  For example,
given \texttt{if(cond) \{*addr=val$_T$;\} else \{*addr=val$_E$;\}}
the transformed code would be 
\texttt{\{*addr = \textbf{\textit{CTSEL}}(cond,val$_T$,*addr);
	*addr = \textbf{\textit{CTSEL}}(cond,*addr,val$_E$);\}}
which has two \emph{CTSEL} instances.
We can remove one or both \emph{CTSEL} instances: 
\begin{itemize}
\item 
If \texttt{(val$_T$==val$_E$)} holds, we merge the two \emph{Store}
operations into one \emph{Store}: \texttt{*addr = val$_T$}.
\item
Otherwise, we use 
\texttt{*addr =  \textbf{\textit{CTSEL}}(cond,val$_T$,val$_E$)}.
\end{itemize}
In the first case, all \emph{CTSEL} instances are avoided.  Even in
the second case, the number of \emph{CTSEL} instances is reduced by
half.

\section{Mitigating Lookup-Table Accesses}
\label{sec:mitigation2}

In this section, we present our method for mitigating lookup-table
accesses that may lead to cache-timing leaks.  In cryptographic
software, such leaks are often due to dependencies between indices
used to access S-Boxes and the secret data.  However, before delving
into the details of our method, we perform a theoretical analysis of
the runtime overhead of various alternatives, including even those
designed against the \emph{more-capable} attackers.

\subsection{Mitigation Granularity and Overhead}

We focus primarily on \emph{less-capable} attackers who only observe
the \emph{total execution time} of the victim's program.  Under this
threat model, we develop optimizations to take advantage of the cache
structure and unique characteristics of the software being protected.
Our mitigation, illustrated by the example in
Figure~\ref{fig:TLUfix4}, can be significantly more efficient than the
approaches illustrated in Figure~\ref{fig:TLUfix2}.


%
In contrast, the \emph{Byte-access-aware} threat model allows
attackers to observe timing characteristics of each instruction in the
victim's program, which means mitigation would have to be applied to
every LUT access to make sure there is no timing difference
(Figure~\ref{fig:TLUfix}).
%

%
The \emph{Line-access-aware} threat model allows attackers to see the
difference between memory locations mapped to different cache lines.
Thus, mitigation only needs to touch all cache lines associated with
the table (Figure~\ref{fig:TLUfix2}).

%
Let $\pi$ be a path in $P$ and $\tau(\pi)$ be its execution time.  Let
$\tau_{max}$ be the maximum value of $\tau(\pi)$ for all possible
$\pi$ in $P$.  For our \emph{Total-time-aware} threat model, the ideal
mitigation would be a program $P'$ whose execution time along all
paths matches $\tau_{max}$.  In this case, we say mitigation
has \emph{no additional} overhead.  We quantify the overhead of other
approaches by comparing to $\tau_{max}$.

Table~\ref{tab:overhead} summarizes the comparison.  Let $N$ be the table
size, $CLS$ be the cache line size, and $M = \lceil N/CLS \rceil$ be
the number of cache lines needed.
Let $K$ be the number of times table elements are accessed. 
Without loss of generality, we assume each element occupies one byte.
In the best case where all $K$ accesses are mapped to the same cache
line, there will be 1 miss followed by $K-1$ hits.
In the worst case ($\tau_{max}$) where the $K$ accesses are scattered
in $M$ cache lines, there will be $M$ misses followed by $K-M$ hits.

\begin{table}
\vspace{1ex}
	\centering
	\caption{Overhead comparison:  
                 $N$ is the table size; 
                 $M=\lceil N/CLS \rceil$ is the number of cache lines to store the table; 
                 $K$ is the number of times table elements are accessed.}
	\label{tab:overhead}
	\resizebox{\linewidth}{!}{\footnotesize
	\begin{tabular}{|p{.45\linewidth}|c|c|c|}
			\hline
			Program Version       & \# Accesses & \# Cache Miss &\# Cache Hit \\
			\hline \hline
		\textbf{Original program}     &K               &from M to 1         & from K-M to K-1  \\
			\hline
			Granularity: Byte-access     &K*N             &M              & K*N-M  \\
			\hline
		        Granularity: Line-access     &K*M             &M              & K*M-M  \\
			\hline
                        Granularity: Total-time ($\tau_{max}$) &K               &M              & K-M  \\
			\hline

\textbf{Our Method: opt. w/ cache analysis}       &K+M-1           &M               & K-1  \\
			\hline
	\end{tabular}
}
\end{table}

When mitigating at the granularity of a byte (e.g.,
Figure~\ref{fig:TLUfix}), the total number of accesses in $P'$ is
increased from $K$ to $K*N$.  Since all elements of the table are
touched when any element is read, all $M$ cache lines will be
accessed.  Thus, there are $M$ cache misses followed by $K*N-M$ hits.

When mitigating at the granularity of a line (e.g.,
Figure~\ref{fig:TLUfix2}), the total number of accesses becomes $K*M$.
Since all cache lines are touched, there are $M$ cache misses followed
by $K*M-M$ hits.

Our method, when equipped with static cache analysis based
optimization (Section~\ref{sec:optimization}), further reduces the
overhead: by checking whether the table, once loaded to the cache,
will stay there until all accesses complete.  If we can prove the
table never gets evicted, we only need to load each line once.
Consequently, there will be $M$ misses in the first loop iteration,
followed by $K-1$ hits in the remaining $K-1$ loop iterations.

In all cases, however, the number of cache misses ($M$) matches that
of the ideal mitigation; the differences is only in the number of
cache hits, which increases from $K-M$ to $K*N-M$, $K*M-M$, or $K-1$.
Although these numbers (of hits) may differ significantly, the actual
time difference may not, because a cache hit often takes an order of
magnitude shorter time than a cache miss.

\begin{figure}
\vspace{1ex}
\centering
\begin{minipage}{0.960\linewidth}
\begin{lstlisting}
//mitigation #6: preloading sbox[256] during the first loop iteration
block_0 = block[0];
for (j=block_0 % CLS; j < 256; j+=CLS) {
	sbox_j = sbox[j];
	val = (block_0 == j)? sbox_j : block_0;
}
block[0] = val;
//access to sbox[...] is always a hit
for (i = 1; i < 16; ++i) {
	block[i] = sbox[block[i]];
}
\end{lstlisting}
\end{minipage}

\vspace{-1ex}
\caption{Reduction: preloading only in the first iteration.}
\label{fig:TLUfix4}
\vspace{-1ex}
\end{figure}

\subsection{Static Cache Analysis-based Reduction}
\label{sec:optimization}

We develop a static cache analysis to compute, at any location,
whether a memory element is definitely in the cache.  This MUST-HIT
analysis~\cite{FerdinandW98,FerdinandW99} allows us to decide
if an LUT access needs mitigation.
For example, in \texttt{subCell()} of \texttt{LED\_encrypt.c} that
accesses \texttt{sbox[16]} using \texttt{for(i=0; i<4; i++)
for(j=0;j<4;j++) \{state[i][j]=sbox[state[i][j]];\}}, since the size
of \texttt{sbox} is 16 bytes while a cache line has 64 bytes, all the
elements can be stored in the same cache line.
Therefore, the first loop iteration would have a cache miss while all
subsequent fifteen iterations would be hits---there is no
cache-timing leak that needs mitigation.

There are many other applications where lookup-table accesses result
in MUST-HITs, e.g., block ciphers with multiple encryption or
decryption rounds, each of which accesses the same lookup table.
Instead of mitigating every round, we use our cache analysis to check
if, starting from the second round, mitigation can be skipped.

\vspace{1ex}\noindent\textbf{Abstract Domain.}
We design our static analysis procedure based on the unified framework
of abstract interpretation~\cite{CousotC77,FerdinandW98,FerdinandW99}, which defines a suitable
abstraction of the program's state as well as transfer functions of
all program statements.
There are two reasons for using abstract interpretation.  The first
one is to ensure the analysis can be performed in finite time even
if precise analysis of the program may be undecidable.  The
second one is to summarize the analysis results along all paths and
for all inputs.

Without loss of generality, we assume the cache has full associativity
and a set $L = \{l_1,..., l_N\}$ of cache lines.  The subscript of
$l_i$, where $1\leq i\leq N$, denotes the age: 1 is the youngest, $N$
is the oldest, and $>N$ means the line is outside of the cache. For
ease of presentation, let $l_{\bot}$ be the imaginary line outside of
the cache.  Thus, $L^* = L \cup \{l_{\bot}\}$ is the extended set of
cache lines.

Let $V = \{v_1,...,v_n\}$ be the set of program variables, each of
which is mapped to a subset $L_v \subseteq L^*$ of cache lines.  The
age of $v\in V$, denoted $Age(v)$, is a set of integers corresponding
to ages (subscripts) of the lines it may reside (along all paths
and for all inputs).  The program's cache state, denoted $S = \langle
Age(v_1), \ldots, Age(v_n) \rangle$, provides the ages of all
variables.

Consider an example program with three variables $x$, $y$ and $z$,
where $x$ is mapped to the first cache line, $y$ may be mapped to the
first two lines (e.g., along two paths) and $z$ may be mapped to Lines
3-5.  Thus, $L_{x} = \{l_1\}$, $L_{y} = \{l_1, l_2\}$, and $L_{z}
= \{l_3, l_4, l_5\}$, and the cache state is
$\langle \{1\}, \{1,2\}, \{3,4,5\} \rangle$.

\vspace{1ex}\noindent\textbf{Transfer Functions.}
The transfer function of each program statement defines how it
transforms the cache state to a new state.
Without loss of generality, we assume the cache uses the
popular \emph{least recent used (LRU)} update policy.  Recall that in
a fully associative cache, a memory block may be mapped to any cache
line; and under LRU, the cache keeps the most recently used memory
blocks while evicting the least recently used blocks.

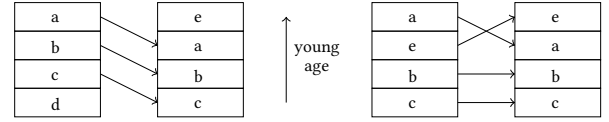
\begin{figure}
\vspace{1ex}
\centering
\scalebox{0.95}{  \begin{tikzpicture}[fill=blue!20,font=\footnotesize]
  \draw (0, 0) rectangle (1.2, 0.4);
  \node[above=5pt, right] at (0.4, 0) {d};
  \draw (0, 0.4) rectangle (1.2, 0.8);
  \node[above=5pt, right] at (0.4, 0.4) {c} ;	
  \draw (0, 0.8) rectangle (1.2, 1.2);
  \node[above=5pt, right] at (0.4, 0.8) {b};
  \draw (0, 1.2) rectangle (1.2, 1.6);
  \node[above=5pt, right] at (0.4, 1.2) {a};
  \draw (2, 0) rectangle (3.2, 0.4);
  \node[above=5pt, right] at (2.4, 0) {c};
  \draw (2, 0.4) rectangle (3.2, 0.8);
  \node[above=5pt, right] at (2.4, 0.4) {b} ;	
  \draw (2, 0.8) rectangle (3.2, 1.2);
  \node[above=5pt, right] at (2.4, 0.8) {a};
  \draw (2, 1.2) rectangle (3.2, 1.6);
  \node[above=5pt, right] at (2.4, 1.2) {e};
  \draw[->] (1.2, 1.4) -- (2, 1.0);
  \draw[->] (1.2, 1.0) -- (2, 0.6);
  \draw[->] (1.2, 0.6) -- (2, 0.2);
  \draw[->] (3.8, 0.2) -- (3.8, 1.4) node [align =center,pos = 0.5,right] {young\\age};
   \draw (5, 0) rectangle (6.2, 0.4);
   \node[above=5pt, right] at (5.4, 0) {c};
   \draw (5, 0.4) rectangle (6.2, 0.8);
   \node[above=5pt, right] at (5.4, 0.4) {b} ;	
   \draw (5, 0.8) rectangle (6.2, 1.2);
   \node[above=5pt, right] at (5.4, 0.8) {e};
   \draw (5, 1.2) rectangle (6.2, 1.6);
   \node[above=5pt, right] at (5.4, 1.2) {a};
   \draw (7, 0) rectangle (8.2, 0.4);
   \node[above=5pt, right] at (7.4, 0) {c};
   \draw (7, 0.4) rectangle (8.2, 0.8);
   \node[above=5pt, right] at (7.4, 0.4) {b} ;	
   \draw (7, 0.8) rectangle (8.2, 1.2);
   \node[above=5pt, right] at (7.4, 0.8) {a};
   \draw (7, 1.2) rectangle (8.2, 1.6);
   \node[above=5pt, right] at (7.4, 1.2) {e};
   \draw[->] (6.2, 1.4) -- (7, 1.0);
   \draw[->] (6.2, 1.0) -- (7, 1.4);
   \draw[->] (6.2, 0.6) -- (7, 0.6);
   \draw[->] (6.2, 0.2) -- (7, 0.2);

  \end{tikzpicture}
 }

\vspace{-1ex}
\caption{Update of the cache: two examples for a fully associative cache, with the LRU update policy.}
\label{fig:cacheupdate}
\vspace{-1ex}
\end{figure}

Figure~\ref{fig:cacheupdate} shows two examples.
On the left-hand side, the initial state, for variables $a$, $b$, $c$,
$d$ and $e$, is $\langle \{1\},\{2\},\{3\},\{4\},\{\bot\} \rangle$.
After accessing $e$, the new state is
$\langle \{2\},\{3\},\{4\},\{\bot\},\{1\} \rangle$.
On the right-hand side, the initial state is
$\langle \{1\},\{3\},\{4\},\{\bot\},\{2\} \rangle$.  After accessing
$e$, the new state is
$\langle \{2\},\{3\},\{4\},\{\bot\},\{1\} \rangle$.
In both cases, the newly accessed $e$ gets the youngest age, while the
ages of other blocks either decrement or remain the same.  Since $d$
is the oldest block (age $5$), it stays outside of the cache.

The transfer function $\mathsf{TFunc}(S,inst)$ models the impact of
instruction $inst$ on the cache state $S$: it returns a new cache
state $S' = \langle Age'(v_1), \ldots, Age'(v_n) \rangle$. 
If $inst$ does not access memory, then $S' = S$.
If $inst$ accesses $v\in P$ in memory, we construct $S'$ as follows:
\begin{itemize}
\item for $v$, set $Age'(v) = \{1\}$ in $S'$;
\item for $u \in V$ such that $\exists a_{u}\in Age(u),a_{v}\in Age(v): a_{u} < a_{v}$ in $S$, replace $a_u$  with $(a_u+1)$ in $Age'(u)$;
\item for any other variable $w \in V$,  $Age'(w) = Age(w)$. 
\end{itemize}
Thus, the function $\mathsf{TFunc}$ models what is illustrated in
Figure~\ref{fig:cacheupdate}.

\vspace{1ex}\noindent\textbf{MUST-HIT Analysis.}
Since our goal is to decide whether a memory block is definitely in
the cache, we compute in $Age(v)$ the upper bound of all possible ages
of $v$, e.g., along all paths and for all inputs.  If this upper bound
is $\leq N$, we know $v$ must be in the cache.

We also define the join ($\sqcup$) operator accordingly; it is needed
to merge states $S$ and $S'$ from different paths.  It is similar to
\emph{set intersection}---in the resulting $S'' = S \sqcup S'$, each
$Age''(v)$ gets the maximum of $Age(v)$ in state $S$ and $Age'(v)$ in
state $S'$. This is because $v\in V$ is definitely in the
cache \emph{only if} it is in the cache according to both states,
i.e., $Age(v) \leq N$ and $Age'(v) \leq N$.

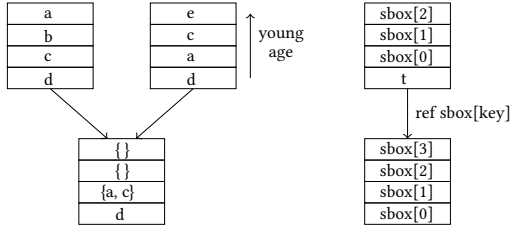
\begin{figure}
\vspace{1ex}
\centering
\scalebox{0.95}{  \begin{tikzpicture}[fill=blue!20,font=\footnotesize]
  \draw (0, 0) rectangle (1.2, 0.3);
  \node[above=4pt, right] at (0.4, 0) {d};
  \draw (0, 0.3) rectangle (1.2, 0.6);
  \node[above=4pt, right] at (0.4, 0.3) {c} ;	
  \draw (0, 0.6) rectangle (1.2, 0.9);
  \node[above=4pt, right] at (0.4, 0.6) {b};
  \draw (0, 0.9) rectangle (1.2, 1.2);
  \node[above=4pt, right] at (0.4, 0.9) {a};
  \draw (2, 0) rectangle (3.2, 0.3);
  \node[above=4pt, right] at (2.4, 0) {d};
  \draw (2, 0.3) rectangle (3.2, 0.6);
  \node[above=4pt, right] at (2.4, 0.3) {a} ;	
  \draw (2, 0.6) rectangle (3.2, 0.9);
  \node[above=4pt, right] at (2.4, 0.6) {c};
  \draw (2, 0.9) rectangle (3.2, 1.2);
  \node[above=4pt, right] at (2.4, 0.9) {e};
  \draw (1.0, -1) rectangle (2.2, -0.7);
  \node[above=4pt, right] at (1.4, -1) { \{ \}};
  \draw (1, -1.3) rectangle (2.2, -1);
  \node[above=4pt, right] at (1.4, -1.3) {\{ \}} ;	
  \draw (1, -1.6) rectangle (2.2, -1.3);
  \node[above=4pt, right] at (1.2, -1.6) {\{a, c\}};
  \draw (1, -1.9) rectangle (2.2, -1.6);
  \node[above=4pt, right] at (1.4, -1.9) {d};
  \draw[->] (0.6, 0) -- (1.4, -0.7);
  \draw[->] (2.6, 0) -- (1.8, -0.7);
  \draw[->] (3.4, 0.15) -- (3.4, 1.05) node [align =center,pos = 0.5,right] {young\\age};
  \draw (5, 0) rectangle (6.2, 0.3);
  \node[above=4pt, right] at (5.4, 0) {t};
  \draw (5, 0.3) rectangle (6.2, 0.6);
  \node[above=4pt, right] at (5.1, 0.3) {sbox[0]};	
  \draw (5, 0.6) rectangle (6.2, 0.9);
  \node[above=4pt, right] at (5.1, 0.6) {sbox[1]};
  \draw (5, 0.9) rectangle (6.2, 1.2);
  \node[above=4pt, right] at (5.1, 0.9) {sbox[2]};
  \draw (5, -1) rectangle (6.2, -0.7);
  \node[above=4pt, right] at (5.1, -1) {sbox[3]};
  \draw (5, -1.3) rectangle (6.2, -1);
  \node[above=4pt, right] at (5.1, -1.3) {sbox[2]} ;	
  \draw (5, -1.6) rectangle (6.2, -1.3);
  \node[above=4pt, right] at (5.1, -1.6) {sbox[1]};
  \draw (5, -1.9) rectangle (6.2, -1.6);
  \node[above=4pt, right] at (5.1, -1.9) {sbox[0]};
  \draw[->] (5.6, 0) -- (5.6, -0.7) node [align =center,pos = 0.5,right] {ref sbox[key]};;

  \end{tikzpicture}
 }

\vspace{-1ex}
\caption{Update of the abstract cache state: (1) on the left-hand side, join at the merge point of two paths; and (2) on the right-hand side, a non-deterministic $key$ for memory access.}
\label{fig:cachejoin}
\vspace{-2ex}
\end{figure}

Consider the left example in Figure~\ref{fig:cachejoin}, where the
ages of $a$ are 1 and 3 before reaching the merge point, and the ages
of $c$ are 3 and 2.  After joining the two cache states, the ages of
$a$ and $c$ become 3, and the age of $d$ remains 4.  The ages of $b$
and $e$ become $\bot$ because, in at least one of the two states, they
are outside of the cache.

Now, consider the right-hand-side example in
Figure~\ref{fig:cachejoin}, where $sbox$ has four elements in total.
In the original state, the first three elements are in the cache
whereas $sbox[3]$ is outside.  After accessing $sbox[key]$, where the
value of $key$ cannot be statically determined, we have to assume the
worst case.  In our MUST-HIT analysis, the worst case means $key$ may
be any index ranging from 0 to 3.  To be safe, we assume $sbox[key]$
is mapped to the oldest element $sbox[3]$.  Thus, the new state has
$sbox[3]$ in the first line while the ages of all other elements are
decremented.

\vspace{1ex}\noindent\textbf{Correctness and Termination.}
Our analysis is a conservative approximation of the actual cache
behavior.  For example, when it says a variable has age 2, its actual
age must not be older than 2.  Therefore, when it says the variable is
in the cache, it is guaranteed to be in the cache, i.e., our analysis is
sound; however, it is not (meant to be) complete in finding all
MUST-HIT cases -- insisting on being both sound and complete could
make the problem undecidable.
In contrast, by ensuring the abstract domain is finite (with finitely
many lines in $L$ and variables in $V$) and both $\mathsf{TFunc}$ and
$\sqcup$ are monotonic, we guarantee that our analysis terminates.

\vspace{1ex}\noindent\textbf{Handling Loops.}
One advantage of abstract interpretation~\cite{CousotC77,FerdinandW98,FerdinandW98} is the
capability of handling loops: for each \emph{back edge} in the CFG,
cache states from all incoming edges are merged using the join
($\sqcup$) operator.
Nevertheless, loops in cryptographic software have unique
characteristics.  For example, most of them have fixed loop bounds,
and many are in functions that are invoked in multiple
encryption/decryption rounds.
Thus, memory accesses often cause cache misses in the first loop
iteration of the first function invocation, but hits subsequently.
Such \emph{first-miss} followed by \emph{always hit}, however,
cannot be directly classified as a MUST-HIT.

To exploit the aforementioned characteristics, we perform a code
transformation prior to our analysis: unrolling the first iteration
out of the loop while keeping the remaining iterations.  For
example, \texttt{for(i=0;i<16;++i) \{block[i]=...\}}
become \texttt{\{block[0]=...\} for(i=1;i<16;++i) \{block[i]=...\}}.
As soon as accesses in the first iteration are mitigated, e.g., as in
Figure~\ref{fig:TLUfix4}, all subsequent loop iterations will result
in MUST-HITs, meaning we can skip the mitigation and avoid the runtime
overhead.
Our experiments on a large number of real applications show that the
cache behaviors of many loops can be exploited in this manner.

\ignore{

We now summarize our analysis using Datalog rules, where \textsc{MR($v$, $k$)} indicates a memory reference on
variable $v$ on offset $k$, $l_{v,k}$ is the cache line mapped to $v$
at offset $k$. 

\vspace{1ex}
{\footnotesize
\begin{itemize}
\item
\noindent
\textsc{MR($v$, $k$) $\longrightarrow$ $age$($l_{v,k}$) = 0}
\item
\noindent
\textsc{MR($v$, $k$) $\&$ $\forall l' \neq l_{v,k}, age(l') \leq age$($l_{v,k}$)$ \longrightarrow$ $age$($l'$) +=1}
\item
\noindent
\textsc{MR($v$, $?$) $\&$ $\exists l_1 \in L_v, \forall l_2 \in L_v, age(l_1) \geq age(l_2)$ $\longrightarrow$ $age$($l_1$) = 0}
\item
\noindent
\textsc{Next($i_1$,$j$) $\&$... $\&$ Next($i_n$,$j$) $\longrightarrow$ $age$($l$)$_j$ = Max($age$($l$)$_{i1}$,..., $age$($l$)$_{in}$) }
\item
\noindent
\textsc{MR(\_,$v$, $k$) $\&$ $age$($l_{v,k}$) $<$ N $\longrightarrow$ Hit(\_,$v$)}
\end{itemize}
}
\vspace{1ex}

The first two rules represent the update on cache state of a memory
reference instruction with specific address. The 3rd rule implement
the update process of non-deterministic memory access as we explained
in Section~\ref{sec:non-deter}. The next rule denotes the cache state
update along the control flow (previously explained using
Figure~\ref{fig:cachejoin}) and the last rule is used to determine if
a memory access would result in cache hit or not.

}

\section{Experiments}
\label{sec:experiment}


We have implemented our method in a tool named \Name{}, which takes
LLVM bit-code as input and returns leak-free bit-code as output.  The
new bit-code may be compiled to machine code to run on any platform
(e.g., x86 and ARM) using standard tools or simulated by GEM5 to
obtain timing statistics.

\ignore{
Our detection of potential timing leaks is accomplished by three
LLVM-based analysis passes:
a sensitivity analysis that reads in the list of secret variables and
propagates the sensitivity attribute to other variables via control-
and data-dependencies;
a static cache analysis invoked on demand to decide whether a store or
load instruction definitely results in a cache hit;
and a leak detection pass leveraging results of the above two
analyses to identify instructions that may cause timing leaks.

Our mitigation is accomplished by two LLVM-based transformation
passes.
The first pass aims to replace sensitive conditional statements with
functionally-equivalent but time-invariant assignments.
The second pass aims to mitigate accesses to sensitive look-up tables
that, depending on the value of the index, may lead to different cache
behaviors.
}


We conducted experiments on C/C++ programs that implement well-known
cryptographic algorithms by compiling them to bit-code using
Clang/LLVM.
Table~\ref{tbl:bench} shows the benchmark statistics.  In total, there
are 19,708 lines of code from libraries including
a real-time Linux kernel (Chronos~\cite{Chronos}), 
a lightweight cryptographic library (FELICS~\cite{FELICS}),
a system for performance evaluation of cryptographic primitives
(SuperCop~\cite{Supercop}),
the Botan cryptographic library~\cite{Botan},
three textbook implementations of cryptographic
algorithms~\cite{AppliedCryp}, and
the GNU \texttt{Libgcrypt} library~\cite{libg}.
Columns~1 and 2 show the benchmark name and source.  Column~3 shows
the number of lines of code (LoC). Columns~4 and 5 show the number of
conditional jumps (\# IF) and the number of lookup tables (\# LUT).
The last two columns show more details of these lookup tables,
including the total, minimum, and maximum table sizes.

\begin{table}[t!]
\vspace{1ex}

\caption{Benchmark statistics.}
\label{tbl:bench}
\centering
\scalebox{0.64}{
\begin{tabular}{|l|l|r|c|c|rc|}
\hline
Name       &Description                                         &\# LoC &\# IF &\# LUT    &\multicolumn{2}{c|}{LUT size in Bytes}\\
\cline{6-7}
           &                                                    &       &    &    &total  & (min, max) \\ 	
\hline\hline
aes        &AES in Chronos~\cite{Chronos}                       &1,379  &3   &5   &16,424 &(40, 4096)  \\
des        &DES in Chronos~\cite{Chronos}                       &874    &2   &11  &6,656  &(256, 4096) \\
des3       &DES-EDE3 in Chronos~\cite{Chronos}                  &904    &2   &11  &6,656  &(256, 4096) \\
anubis     &Anubis in Chronos~\cite{Chronos}                    &723    &1   &7   &6,220  &(76, 1024)  \\
cast5      &Cast5 cipher (rfc2144) in Chronos~\cite{Chronos}    &799    &0   &8   &8,192  &(1024, 1024)\\
cast6      &Cast6 cipher (rfc2612) in Chronos~\cite{Chronos}    &518    &0   &6   &4,896  &(32, 1024)  \\ 
fcrypt     &FCrypt encryption in Chronos~\cite{Chronos}         &401    &0   &4   &4,096  &(1024, 1024)\\
khazad     &Khazad algorithm in Chronos~\cite{Chronos}          &841    &0   &9   &16,456 &(72, 2048)  \\
\hline
LBlock     &LBlock cipher from Felics~\cite{FELICS}             &1,005  &0   &10  &160    &(16,16)     \\
Piccolo    &Piccolo cipher from Felics~\cite{FELICS}            &243    &2   &4   &148    &(16,100)     \\
PRESENT    &PRESENT cipher from Felics~\cite{FELICS}            &183    &0   &33  &2,064  &(15,64)     \\
TWINE      &TWINE  cipher from Felics~\cite{FELICS}             &249    &0   &3   &67     &(16,35)     \\
\hline
aes        &AES in SuperCop~\cite{Supercop}                     &1099   &4   &10  &8,488  &(40, 1024)  \\
cast       &CAST in SuperCop~\cite{Supercop}                    &942    &5   &8   &16,384 &(2048, 2048)\\
\hline
aes\_key   &AES key\_schedule in Botan~\cite{Botan}             &502    &3   &4   &8,704  &(256,4096)  \\
cast128    &cast 128-bit in Botan~\cite{Botan}                  &617    &2   &8   &8,192  &(1024,1024) \\
des        &des cipher in Botan~\cite{Botan}                    &835    &1   &12  &10,240 &(1024,2048) \\
kasumi     &kasumi cipher  in Botan~\cite{Botan}                &275    &2   &2   &1,152  &(128,1024)  \\
seed       &seed cipher in Botan~\cite{Botan}                   &352    &0   &5   &4,160  &(64,1024)   \\
twofish    &twofish cipher in Botan~\cite{Botan}                &770    &18  &9   &5,150  &(32,1024)   \\
\hline
3way       &3way cipher reference~\cite{AppliedCryp}            &177    &10  &0   &0      &(0,0)       \\
des        &des cipher reference~\cite{AppliedCryp}             &463    &16  &14  &2,302  &(16,512)    \\
loki91     &loki cipher reference~\cite{AppliedCryp}            &231    &10  &1   &32     &(32,32)     \\
\hline
camellia  &camellia cipher in Libgcrypt~\cite{libg}             &1453   &0   &4   &4,096  &(1024,1024) \\
des       &des cipher in Libgcrypt~\cite{libg}                  &1486   &2   &13  &2,724  &(16,2048) \\
seed      &seed cipher in Libgcrypt~\cite{libg}                 &488    &3   &5   &4,160  &(64,1024) \\
twofish   &twofish cipher in Libgcrypt~\cite{libg}              &1899   &1   &6   &6,380  &(256,4096) \\
\hline
\end{tabular}
}
\vspace{-2ex}
\end{table}

Our experiments aimed to answer three research questions:
(1) Is our method effective in mitigating instruction-
and cache-timing leaks?
(2) Is our method efficient in handling real applications?
(3) Is the overhead of the mitigated code,  in terms of code size
and run time, low enough for practical use?

\subsection{Results: Leak Detection}

Table~\ref{tbl:detection} shows the results of applying our leak
detection technique, where Columns~1-4 show the name of the benchmark
together with the number of conditional jumps (\# IF), lookup tables
(\# LUT), and accesses to table elements (\# LUT-access),
respectively.  Columns~5-7 show the number of \emph{sensitive}
conditional jumps, lookup tables, and accesses, respectively.  Thus,
non-zero in the sensitive \texttt{\#IF} column means there is
instruction-timing leakage, and non-zero in the
sensitive \texttt{\#LUT-access} means there is cache-timing leakage.
We omit the time taken by our static analysis since it is negligible:
in all cases the analysis completed in a few seconds.

\begin{table}[t!]
\vspace{1ex}

\caption{Results of conducting static leak detection.}
\label{tbl:detection}
\centering
\scalebox{.65}{
\begin{tabular}{|l|ccr|ccr|}
\hline
            
Name          &\multicolumn{3}{c|}{Total}  &\multicolumn{3}{c|}{Sensitive (leaky)} \\
\cline{2-7}
              &~~~~~\# IF~~~~~    &~~~\# LUT~~~   &\# LUT-access    &~~~~~\# IF~~~~~    &~~~\# LUT~~~   &\# LUT-access    \\ 
\hline\hline
aes           &3       &5       &424       &0       &4       &416     \\
des           &2       &11      &640       &0       &11      &640     \\
des3          &2       &11      &1,152     &0       &11      &1,152    \\
anubis        &1       &7       &871       &0       &6       &868     \\
cast5         &0       &8       &448       &0       &8       &448     \\
cast6         &0       &6       &448       &0       &4       &384    \\
fcrypt        &0       &4       &128       &0       &4       &128    \\
khazad        &0       &9       &240       &0       &8       &248    \\
\hline
*LBlock        &0       &10      &320      &0       &0       &0          \\
*Piccolo       &2       &4       &121      &0       &0       &0          \\
*PRESENT~~~~~~~&0       &33      &1,056    &0       &0       &0          \\
*TWINE         &0       &3       &156      &0       &0       &0          \\
\hline
aes           &4       &10      &706       &0       &9       &696     \\
cast          &5       &8       &448       &0       &8       &448    \\
\hline
aes\_key      &3       &4       &784       &0       &2       &184     \\
cast128       &2       &8       &448       &0       &8       &448    \\
des           &1       &12      &264       &0       &8       &256    \\
kasumi        &2       &2       &192       &0       &2       &192     \\
seed          &0       &5       &576       &0       &4       &512    \\
twofish       &18      &9       &2,576     &16      &8       &2,512    \\
\hline
3way          &10      &0       &0         &3       &0       &0      \\
des           &16      &14      &456       &2       &8       &128    \\
loki91        &10      &1       &512       &4       &0       &0     \\
\hline
camellia      &0       &4       &32        &0       &4       &32        \\
des           &2       &13      &231       &0       &8       &128    \\
seed          &3       &5       &518       &0       &4       &200     \\
twofish       &1       &6       &8,751     &0       &5       &2,576    \\
\hline
\end{tabular}
}
\vspace{-2ex}
\end{table}

Although conditional statements (\#IF) exist, few are sensitive.
Indeed, only \texttt{twofish} from Botan\cite{Botan} and three old
textbook implementations (\texttt{3way}, \texttt{des},
and \texttt{loki91}) have leaks of this type.
%
In contrast, many lookup tables are sensitive due to cache.
This result was obtained using fully associative LRU cache with 512
cache lines, 64 bytes per line, and thus 32 Kilobytes in total.


Some benchmarks, e.g., \texttt{aes\_key} from Botan~\cite{Botan},
already preload lookup tables; however, our analysis still reports
timing leakage, as shown in Figure~\ref{fig:casestudy},
where \texttt{XEK} is key-related and used to access an array in the
second for-loop.  Although the table named \texttt{TD} is computed at
the run time (thus capable of avoiding \texttt{flush+reload} attack)
and all other tables are preloaded before accesses, it forgot to
preload \texttt{SE[256]}, which caused the cache-timing leak.
%
%
%
%

\begin{figure}
\vspace{1ex}

	\centering
	\begin{minipage}{0.960\linewidth}
		\begin{lstlisting}
const uint8_t SE[256] = {0x63, 0x7C, 0x77, 0x7B,...};
void aes_key_schedule(const uint8_t key[], size_t length,
	std::vector<uint32_t>& EK, std::vector<uint32_t>& DK,
	std::vector<uint8_t>& ME, std::vector<uint8_t>& MD)  {
	static const uint32_t RC[10] = {0x01000000, 0x02000000,...};
	std::vector<uint32_t> XEK(48), XDK(48);
	const std::vector<uint32_t>& TD = AES_TD();
	for(size_t i = 0; i != 4; ++i)
		XEK[i] = load_be<uint32_t>(key, i);
	for(size_t i = 4; i < 44; i += 4) {
		XEK[i] = XEK[i-4] ^ RC[(i-4)/4] ^
             make_uint32(SE[get_byte(1, XEK[i-1])],
									SE[get_byte(2, XEK[i-1])],
									SE[get_byte(3, XEK[i-1])],
									SE[get_byte(0, XEK[i-1])]);
		...
	}
	...
}
\end{lstlisting}
	\end{minipage}

\vspace{-2ex}
	\caption{Reduction: preloading only in the first iteration.}
	\label{fig:casestudy}
\vspace{-2ex}
\end{figure}

\subsection{Results: Leak Mitigation}

To evaluate whether our method can robustly handle real applications,
we collected results of applying our mitigation procedure to each
benchmark. 
%
%
Table~\ref{tbl:result1} shows the results. 
Specifically, Columns~2-5 show the result of our mitigation without
cache analysis-based optimization, while Columns~6-9 show the result
with the optimization.  In each case, we report the number of LUT
accesses actually mitigated, the time taken to complete the
mitigation, the increase in program size, and the increase in runtime
overhead.
For \texttt{anubis}, in particular, our cache analysis showed that
only 10 out of the 868 sensitive LUT accesses needed mitigation; as a
result, optimization reduced both the program's size (from 9.08x to
1.10x) and its execution time (from 6.90x to 1.07x).

\begin{table}[t!]

\caption{Results of leak mitigation.
Runtime overhead is based on average of 1000 simulations with random keys.}

\label{tbl:result1}
\centering
\scalebox{.65}{
\begin{tabular}{|l|rrrr|rrrr|}
\hline
\multirow{2}{*}{Name} 		&  
\multicolumn{4}{c|}{Mitigation w/o opt}		&
\multicolumn{4}{c|}{Mitigation w/ opt}		\\
\cline{2-9} 
             &\# LUT-a &Time(s)  &Prog-size &Ex-time &\# LUT-a &Time(s)  &Prog-size  &Ex-time\\ \hline\hline
			aes          &416   &0.61   & 5.40x	&2.70x  &20   &0.28  &1.22x	&1.11x  \\ 
			des          &640   &1.17   &19.50x	&5.68x  &22   &0.13  &1.23x	&1.07x  \\   
			des3         &1,152 &1.80   &12.90x    &12.40x  &22   &0.46  &1.13x	&1.07x  \\   
			anubis       &868   &3.12   & 9.08x	&6.90x  &10   &0.75  &1.10x	&1.07x  \\
			cast5        &448   &0.79   & 7.24x	&3.84x  &12   &0.22  &1.18x	&1.07x  \\
			cast6        &384   &0.72   & 7.35x	&3.48x  &12   &0.25  &1.16x	&1.08x  \\
			fcrypt       &128   &0.07   & 5.70x	&1.59x  &8    &0.03  &1.34x	&1.05x  \\
			khazad       &248   &0.45   & 8.60x	&4.94x  &16   &0.07  &1.49x	&1.35x  \\\hline                                                            
			aes          &696   &0.96   & 9.52x	&2.39x  &18   &0.22  &1.21x	&1.06x  \\
			cast         &448   &1.42   &13.40x	&6.50x  &12   &0.30  &1.35x	&1.20x  \\\hline                                                            
			aes\_key     &184   &0.27   & 1.35x	&1.19x  &1    &0.23  &1.00x	&1.00x  \\
			cast128      &448   &0.42   & 3.62x	&2.48x  &12   &0.10  &1.09x	&1.06x  \\
			des          &256   &0.21   & 3.69x	&1.86x  &16   &0.06  &1.17x	&1.08x  \\
			kasumi       &192   &0.18   & 2.27x	&1.37x  &4    &0.11  &1.03x	&1.01x  \\
			seed         &512   &0.57   & 6.18x	&1.94x  &12   &0.15  &1.12x	&1.03x  \\
			twofish      &2,512 &29.70  & 5.69x	&4.77x  &8    &10.6  &1.02x	&1.03x  \\\hline                                                            
			3way         &0     &0.01   & 1.01x	&1.03x  &0    &0.01  &1.01x	&1.03x  \\
			des          &128   &0.05   & 2.21x	&1.22x  &8    &0.03  &1.09x	&1.11x  \\
			loki91       &0     &0.01   & 1.01x	&2.83x  &0    &0.01  &1.01x	&2.83x  \\\hline
			camellia     &32    &0.04   & 2.21x	&1.35x  &4    &0.03  &1.20x	&1.09x  \\
			des          &128   &0.06   & 2.30x	&1.20x  &8    &0.03  &1.10x	&1.02x  \\
			seed         &200   &0.01   & 1.38x	&1.36x  &8    &0.01  &1.20x	&1.18x  \\
			twofish      &2,576 &32.40  & 6.85x	&6.59x  &136 &11.90  &1.41x	&1.46x  \\\hline                                     
\end{tabular}		
	}
\end{table}

\begin{wrapfigure}{R}{0.42\linewidth}
\vspace{-2ex}
\begin{minipage}{.8\linewidth}
\hspace{1mm}
    \begin{tikzpicture}[trim axis left]
      \begin{axis}[
		scale only axis,
		width=.95\textwidth,
		xmin=1,
		xmax=3,
		ymin=1,
		ymax=3,
		yticklabel style={align=right,inner sep=0pt,xshift=-0.1cm},
                xlabel={\scriptsize Ex-time (with CMOV)},
                ylabel={\scriptsize Ex-time (with Bitwise)},
		grid=both,
		legend style={legend pos=north west
			, font=\tiny
			, mark options={scale=0.5}
			, inner sep=0.1pt
			, row sep=-0.12cm}, 
                every axis/.append style={font=\tiny}
		]
		\addplot+[only marks, mark size=1.25pt] table {scatter_bitwise_vs_cmov_time_overhead.dat};
		\draw[ultra thin]  (axis cs:1,1) -- (axis cs:3,3);
	\end{axis}
      \end{tikzpicture}
\end{minipage}

\vspace{-2ex}
\caption{Comparing \emph{CTSEL} implementations.}
\label{fig:scatter1}
\vspace{-4ex}
\end{wrapfigure}
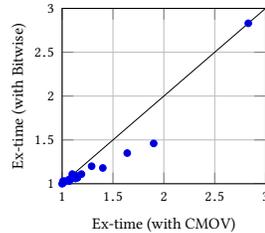

We also compared the execution time with generic (bitwise) versus
optimized (\emph{CMOV}) implementations of \emph{CTSEL(c,t,e)}.
Figure~\ref{fig:scatter1} shows the result in a scatter plot, where
points below the diagonal line are cases where the optimized
implementation is faster.

\subsection{Results: GEM5 Simulation}

Although our analysis is conservative in that the mitigated code is
guaranteed to be leak-free, it is still useful to conduct GEM5
simulations, for two reasons.  First, it confirms our analysis
reflects the reality: the reported leaks are real.  Second, it
demonstrates that, after mitigation, leaks are indeed
eliminated.

Table~\ref{tbl:result2} shows our results.  For each benchmark, we ran
the machine code compiled for x86 on GEM5 using two manually crafted
inputs (e.g., cryptographic keys) capable of showing the timing
variation.
Columns~2-5 show the results of the original program, including the
number of CPU cycles taken to execute it under the two inputs, as well
as the number of cache misses.  Columns~6-9 show the results on the
mitigated program versions.

The results show the execution time of the original program indeed
varies, indicating there are leaks.  But it becomes constant after
mitigation, indicating leaks are removed. 
The one exception is \emph{aes\_keys}: we were not able to manually
craft the inputs under which leak is demonstrable on GEM5.  Since the
input space is large, manually crafting such inputs is not always
easy: symbolic execution tools~\cite{GuoKWYG15,GuoKW16,GuoWW17,GuoWW18} may help generate such leak-manifesting
input pairs --- we will consider it for future work.

\begin{table}[t!]
\vspace{1ex}

\caption{Results of GEM5 simulation with 2 random inputs.}
\label{tbl:result2}
\centering
\scalebox{.64}{
\begin{tabular}{|l|rr|rr|cc|cc|}
        \hline
	\multirow{2}{*}{Name} 		&  
	\multicolumn{4}{c|}{Before Mitigation} &		
	\multicolumn{2}{c|}{Mitigation w/o opt}	&
	\multicolumn{2}{c|}{Mitigation w/ opt}	\\
	\cline{2-9} 
                       &\multicolumn{2}{c|}{\# CPU cycle (in$_1$,in$_2$)}
                                            &\multicolumn{2}{c|}{\# Miss (in$_1$,in$_2$)}
                                                           &\# CPU cycle   &\# Miss &\# CPU cycle &\# Miss \\ \hline\hline
	aes            &100,554 &101,496  &261 &269  &  204,260    &303  &  112,004   &303  \\
	des            & 95,630	& 90,394  &254 &211  &  346,170    &280  &  100,694   &280  \\
	des3           &118,362 &111,610  &271 &211  &  865,656    &280  &  124,176   &280  \\
	anubis         &128,602 &127,514  &276 &275  &  512,452    &276  &  134,606   &276  \\
	cast5          &102,426 &102,070  &282 &279  &  266,156    &304  &  108,068   &304  \\
	cast6          & 96,992	& 97,492  &238 &245  &  233,774    &245  &  100,914   &245  \\
	fcrypt         & 84,616	& 83,198  &224 &218  &  114,576    &240  &   88,236   &240  \\
	khazad         &101,844 &100,724  &332 &322  &  366,756    &432  &  130,682   &432  \\\hline
	aes            & 89,968	& 90,160  &234 &235  &  174,904    &240  &   94,364   &240  \\
	cast           &117,936 &117,544  &345 &342  &  520,336    &436  &  136,052   &435  \\\hline
	aes\_key*      &243,256 &243,256  &329 &329  &  254,262    &329  &  245,584   &328  \\
	cast128        &161,954 &161,694  &298 &296  &  305,514    &321  &  167,626   &321  \\
	des            &118,848 &119,038  &269 &270  &  182,830    &317  &  127,374   &316  \\
	kasumi         &113,362 &113,654  &204 &206  &  137,914    &206  &  115,060   &206  \\
	seed           &106,518 &106,364  &239 &238  &  165,546    &249  &  110,486   &249  \\
	twofish        &309,160 &299,956  &336 &334  &1,060,832    &340  &  315,018   &339  \\\hline
	3way           & 87,834	& 87,444  &181 &181  &   90,844    &182  &   90,844   &182  \\  
	des            &152,808 &147,344  &224 &222  &  181,074    &225  &  168,938   &225  \\
	loki91         &768,064 &768,296  &181 &181  &2,170,626    &181  &2,170,626   &181  \\\hline
	camellia       & 84,208	& 84,020  &205 &203  &  102,100    &244  &   91,180   &244  \\
	des            &100,396 &100,100  &212 &211  &  112,992    &213  &  100,500   &213  \\
	seed           & 83,256	& 83,372  &228 &230  &  107,318    &240  &   96,266   &239  \\
	twofish        &230,838 &229,948  &334 &327  &  982,258    &338  &  295,268   &338  \\	
	\hline
	\end{tabular}
	}
\vspace{-2ex}
\end{table}

\subsection{Threats to Validity}

%
%
First, our mitigation is software based; as such, we do not address
leaks exploitable only by probing the hardware such as instruction
pipelines and data buses.
We focus on the \emph{total-time-aware} threat model: although
extensions to handle other threat models are possible (e.g.,
multi-core and multi-level cache), we consider them as future work.
It is possible that timing characteristics of the machine code may
differ from those of the LLVM bit-code, but we have taken efforts in
making sure machine code produced by our tool does not deviate from
the mitigated bit-code.  For example, we always align sensitive lookup
tables to cache line boundaries, and we implement \emph{CTSEL} as an
intrinsic function to ensure constant-time execution. We also use
GEM5 simulation to confirm that machine code produced by our tool is
indeed free of timing leaks.

\section{Related Work}
\label{sec:related}

Kocher~\cite{Kocher96} is perhaps the first to publicly demonstrate the
feasibility of timing side-channel attacks in embedded systems.  Since
then, timing attacks have been demonstrated on many 
platforms~\cite{AlFardanP13,BrumleyB05,CockGMH14,OsvikST06,GrabherGP07,BangAPPB16,PhanBPMB17,JiangFK16,WangWLZW17}.
For example,
Brumley et al.~\cite{BrumleyB05} showed timing attacks
could be carried out remotely through a computer network.
Cock et al.~\cite{CockGMH14} found timing side channels in the seL4
microkernel and then performed a quantitative evaluation.


Noninterference
properties~\cite{AlFardanP13,HedinS05,BartheRW06,KopfM07,PierroHW11}
have also been formulated to characterize side-channel leaks.
To quantify these leaks, Millen~\cite{Millen87} used Shannon's channel
capacity~\cite{Shannon48} to model the correlation between sensitive
data and timing observations.  Other approaches, including
min-entropy~\cite{Smith09} and $g$-leakage~\cite{AlvimCMMPS14}, were
also developed.  Backes and K\"opf~\cite{BackesK08} developed an
information-theoretic model for quantifying the leaked 
information. K\"opf and Smith~\cite{KopfS10} also proposed a technique for
bounding the leakage in blinded cryptographic algorithms.
%


Prior countermeasures for timing leaks focused primarily on
conditional branches, e.g., type-driven cross-copying~\cite{Agat00}.
Molnar et al.~\cite{MolnarPSW05} introduced, along the \emph{program
counter} model, a method for merging branches.  K\"opf and
Mantel~\cite{KopfM07} proposed a unification-based technique
encompassing the previous two methods.  Independently, Barthe et
al.~\cite{BartheRW06} proposed a transactional branching technique
that leverages commit/abort operations.  Coppens et
al.~\cite{CoppensVBS09} developed a compiler backend for removing such
leaks on x86 processors.
However, Mantel and Starostin~\cite{MantelS15} recently compared four
of these existing techniques on Java byte-code, and showed that none
was able to eliminate the leaks completely.
%
%
%
Furthermore, these methods did not consider cache-timing leaks.



There are techniques that do not eliminate but hide timing leaks via
randomization or
blinding~\cite{Hu91,CraneHBLF15,Kocher96,KopfD09,AskarovZM10,ZhangAM12,BraunJB15}.
%
%
%
%
%
There are also hardware-based mitigation techniques, which fall into
two categories: resource isolation and timing obfuscation.
Resource isolation~\cite{page2005,liu2016, wang2007} may be realized
by partitioning hardware to two parts (public and private)
and then restrict sensitive data/operations to the private partition.
However, it requires modifications of the CPU which is not
always possible.
Timing obfuscation~\cite{Hu91,vattikonda2011,rane2015} may be achieved
by inserting fixed or random delays, or interfering the measurement of
the system clock.
In addition to being expensive, such techniques do not eliminate
timing channels.
Oblivious RAM~\cite{liu2015,stefanov2013, goldreich1996} is another
technique for removing leakage through the data flows, but requires a
substantial amount of on-chip memory and incurs significant overhead
in the execution time.
%

Beyond timing side channels, there are countermeasure techniques for
mitigating leaks through other side channels including
power~\cite{KocherJJ99,MangardOP07} and faults~\cite{BihamS97}.  Some
of these techniques have been automated in compiler-like
tools~\cite{BayrakRBSI11,MossOPT12,AgostaBP12} whereas others have
leveraged the more sophisticated, SMT solver-based, formal
verification~\cite{EldibWS14,EldibWTS14,ZhangGSW18} and inductive
synthesis techniques~\cite{EldibW14,EldibWW16,WangS17}.
However, none of these compiler or formal methods based techniques was
applied to cache-timing side channels.

\section{Conclusions}
\label{sec:conclusion}

We have presented a method for mitigating side-channel leaks via
program repair. The method was implemented in \Name{}, a tool for
handling cryptographic libraries written in C/C++.  We evaluated it on
real applications and showed the method was scalable and efficient,
while being effective in removing both instruction- and cache-related
timing side channels. Furthermore, the mitigated software code had
only moderate increases in program size and run-time
overhead.

\section*{Acknowledgments}

This work was supported in part by the NSF under grants CNS-1617203 and
CNS-1702824 and ONR under grant N00014-17-1-2896.
%


\clearpage
\newpage
\bibliographystyle{plain}
\bibliography{tsyn,tapsc}

\begin{thebibliography}{10}

\bibitem{Botan}
{\em {Botan: Crypto and TLS for C++11}}.
\newblock \url{https://github.com/randombit/botan/}.

\bibitem{FELICS}
{\em {Fair Evaluation of Lightweight Cryptographic Systems}}.
\newblock \url{https://www.cryptolux.org/index.php/FELICS}.

\bibitem{Libgcrypt}
{\em {Libgcrypt}}.
\newblock \url{https://gnupg.org/software/libgcrypt/index.html}.

\bibitem{libg}
{\em {Libgcrypt}}.
\newblock \url{https://www.gnupg.org/software/libgcrypt/index.html}.

\bibitem{Supercop}
{\em {System for Unified Performance Evaluation Related to Cryptographic
  Operations and Primitives}}.
\newblock \url{https://bench.cr.yp.to/supercop.html}.

\bibitem{LLVM}
{\em {The LLVM Compiler Infrastructure}}.
\newblock \url{ http://llvm.org/}.

\bibitem{Agat00}
Johan Agat.
\newblock Transforming out timing leaks.
\newblock In {\em ACM SIGACT-SIGPLAN Symposium on Principles of Programming
  Languages}, pages 40--53, 2000.

\bibitem{AgostaBP12}
Giovanni Agosta, Alessandro Barenghi, and Gerardo Pelosi.
\newblock A code morphing methodology to automate power analysis
  countermeasures.
\newblock In {\em ACM/IEEE Design Automation Conference}, pages 77--82, 2012.

\bibitem{AlFardanP13}
Nadhem~J. AlFardan and Kenneth~G. Paterson.
\newblock Lucky thirteen: Breaking the {TLS} and {DTLS} record protocols.
\newblock In {\em {IEEE} Symposium on Security and Privacy}, pages 526--540,
  2013.

\bibitem{AlvimCMMPS14}
M{\'{a}}rio~S. Alvim, Konstantinos Chatzikokolakis, Annabelle McIver, Carroll
  Morgan, Catuscia Palamidessi, and Geoffrey Smith.
\newblock Additive and multiplicative notions of leakage, and their capacities.
\newblock In {\em {IEEE} Computer Security Foundations Symposium}, pages
  308--322, 2014.

\bibitem{AntonopoulosGHK17}
Timos Antonopoulos, Paul Gazzillo, Michael Hicks, Eric Koskinen, Tachio
  Terauchi, and Shiyi Wei.
\newblock Decomposition instead of self-composition for proving the absence of
  timing channels.
\newblock In {\em ACM SIGPLAN Conference on Programming Language Design and
  Implementation}, pages 362--375, 2017.

\bibitem{AskarovZM10}
Aslan Askarov, Danfeng Zhang, and Andrew~C. Myers.
\newblock Predictive black-box mitigation of timing channels.
\newblock In {\em {ACM} Conference on Computer and Communications Security},
  pages 297--307, 2010.

\bibitem{AwekeA17}
Zelalem~Birhanu Aweke and Todd~M. Austin.
\newblock {\O}zone: Efficient execution with zero timing leakage for modern
  microarchitectures.
\newblock In {\em {IEEE} International Symposium on Hardware Oriented Security
  and Trust}, page 153, 2017.

\bibitem{BackesK08}
Michael Backes and Boris K{\"{o}}pf.
\newblock Formally bounding the side-channel leakage in unknown-message
  attacks.
\newblock In {\em European Symposium on Research in Computer Security}, pages
  517--532, 2008.

\bibitem{BalatsourasS16}
George Balatsouras and Yannis Smaragdakis.
\newblock Structure-sensitive points-to analysis for {C} and {C++}.
\newblock In {\em International Symposium on Static Analysis}, pages 84--104,
  2016.

\bibitem{BangAPPB16}
Lucas Bang, Abdulbaki Aydin, Quoc{-}Sang Phan, Corina~S. Pasareanu, and Tevfik
  Bultan.
\newblock String analysis for side channels with segmented oracles.
\newblock In {\em ACM SIGSOFT Symposium on Foundations of Software
  Engineering}, pages 193--204, 2016.

\bibitem{BartheRW06}
Gilles Barthe, Tamara Rezk, and Martijn Warnier.
\newblock Preventing timing leaks through transactional branching instructions.
\newblock {\em Electr. Notes Theor. Comput. Sci.}, 153(2):33--55, 2006.

\bibitem{BasuC17}
Tiyash Basu and Sudipta Chattopadhyay.
\newblock Testing cache side-channel leakage.
\newblock In {\em {IEEE} International Conference on Software Testing,
  Verification and Validation Workshops}, pages 51--60, 2017.

\bibitem{BayrakRBSI11}
Ali~Galip Bayrak, Francesco Regazzoni, Philip Brisk, Fran\c{c}ois-Xavier
  Standaert, and Paolo Ienne.
\newblock A first step towards automatic application of power analysis
  countermeasures.
\newblock In {\em ACM/IEEE Design Automation Conference}, pages 230--235, 2011.

\bibitem{BihamS97}
Eli Biham and Adi Shamir.
\newblock Differential fault analysis of secret key cryptosystems.
\newblock In {\em International Cryptology Conference}, pages 513--525, 1997.

\bibitem{GEM5}
Nathan~L. Binkert, Bradford~M. Beckmann, Gabriel Black, Steven~K. Reinhardt,
  Ali~G. Saidi, Arkaprava Basu, Joel Hestness, Derek Hower, Tushar Krishna,
  Somayeh Sardashti, Rathijit Sen, Korey Sewell, Muhammad Shoaib~Bin Altaf,
  Nilay Vaish, Mark~D. Hill, and David~A. Wood.
\newblock The gem5 simulator.
\newblock {\em {SIGARCH} Computer Architecture News}, 39(2):1--7, 2011.

\bibitem{bortz2007}
Andrew Bortz and Dan Boneh.
\newblock Exposing private information by timing web applications.
\newblock In {\em International Conference on World Wide Web}, pages 621--628,
  2007.

\bibitem{BraunJB15}
Benjamin~A. Braun, Suman Jana, and Dan Boneh.
\newblock Robust and efficient elimination of cache and timing side channels.
\newblock {\em CoRR}, abs/1506.00189, 2015.

\bibitem{BrumleyB05}
David Brumley and Dan Boneh.
\newblock Remote timing attacks are practical.
\newblock {\em Computer Networks}, 48(5):701--716, 2005.

\bibitem{Chattopadhyay17}
Sudipta Chattopadhyay.
\newblock Directed automated memory performance testing.
\newblock In {\em International Conference on Tools and Algorithms for
  Construction and Analysis of Systems}, pages 38--55, 2017.

\bibitem{ChenFD17}
Jia Chen, Yu~Feng, and Isil Dillig.
\newblock Precise detection of side-channel vulnerabilities using quantitative
  cartesian hoare logic.
\newblock In {\em Proceedings of the 2017 {ACM} {SIGSAC} Conference on Computer
  and Communications Security}, pages 875--890, 2017.

\bibitem{ChuJM16}
Duc{-}Hiep Chu, Joxan Jaffar, and Rasool Maghareh.
\newblock Precise cache timing analysis via symbolic execution.
\newblock In {\em {IEEE} Real-Time and Embedded Technology and Applications
  Symposium}, pages 293--304, 2016.

\bibitem{CockGMH14}
David Cock, Qian Ge, Toby~C. Murray, and Gernot Heiser.
\newblock The last mile: An empirical study of timing channels on {seL4}.
\newblock In {\em {ACM} {SIGSAC} Conference on Computer and Communications
  Security}, pages 570--581, 2014.

\bibitem{CoppensVBS09}
Bart Coppens, Ingrid Verbauwhede, Koen~De Bosschere, and Bjorn~De Sutter.
\newblock Practical mitigations for timing-based side-channel attacks on modern
  x86 processors.
\newblock In {\em {IEEE} Symposium on Security and Privacy}, pages 45--60,
  2009.

\bibitem{CousotC77}
Patrick Cousot and Radhia Cousot.
\newblock Abstract interpretation: A unified lattice model for static analysis
  of programs by construction or approximation of fixpoints.
\newblock In {\em ACM SIGACT-SIGPLAN Symposium on Principles of Programming
  Languages}, pages 238--252, 1977.

\bibitem{CraneHBLF15}
Stephen Crane, Andrei Homescu, Stefan Brunthaler, Per Larsen, and Michael
  Franz.
\newblock Thwarting cache side-channel attacks through dynamic software
  diversity.
\newblock In {\em Annual Network and Distributed System Security Symposium},
  2015.

\bibitem{Chronos}
Matthew Dellinger, Piyush Garyali, and Binoy Ravindran.
\newblock Chronos linux: a best-effort real-time multiprocessor linux kernel.
\newblock In {\em ACM/IEEE Design Automation Conference}, pages 474--479, 2011.

\bibitem{DinuBGKCP2015}
Daniel Dinu, Yann~Le Corre, Dmitry Khovratovich, Leo Perrin, Johann Grobschadl,
  and Alex Biryukov.
\newblock Triathlon of lightweight block ciphers for the internet of things.
\newblock Cryptology ePrint Archive, Report 2015/209, 2015.

\bibitem{DoychevFKMR13}
Goran Doychev, Dominik Feld, Boris K{\"o}pf, Laurent Mauborgne, and Jan
  Reineke.
\newblock {CacheAudit}: A tool for the static analysis of cache side channels.
\newblock In {\em USENIX Security}, pages 431--446, 2013.

\bibitem{DoychevKMR15}
Goran Doychev, Boris K{\"{o}}pf, Laurent Mauborgne, and Jan Reineke.
\newblock Cacheaudit: {A} tool for the static analysis of cache side channels.
\newblock {\em {ACM} Trans. Inf. Syst. Secur.}, 18(1):4:1--4:32, 2015.

\bibitem{EldibW14}
Hassan Eldib and Chao Wang.
\newblock Synthesis of masking countermeasures against side channel attacks.
\newblock In {\em International Conference on Computer Aided Verification},
  pages 114--130, 2014.

\bibitem{EldibWS14}
Hassan Eldib, Chao Wang, and Patrick Schaumont.
\newblock {SMT}-based verification of software countermeasures against
  side-channel attacks.
\newblock In {\em International Conference on Tools and Algorithms for
  Construction and Analysis of Systems}, pages 62--77, 2014.

\bibitem{EldibWTS14}
Hassan Eldib, Chao Wang, Mostafa Taha, and Patrick Schaumont.
\newblock {QMS}: Evaluating the side-channel resistance of masked software from
  source code.
\newblock In {\em ACM/IEEE Design Automation Conference}, pages 209:1--6, 2014.

\bibitem{EldibWW16}
Hassan Eldib, Meng Wu, and Chao Wang.
\newblock Synthesis of fault-attack countermeasures for cryptographic circuits.
\newblock In {\em International Conference on Computer Aided Verification},
  pages 343--363, 2016.

\bibitem{FerdinandW98}
Christian Ferdinand and Reinhard Wilhelm.
\newblock On predicting data cache behavior for real-time systems.
\newblock In {\em Languages, Compilers, and Tools for Embedded Systems, {ACM}
  {SIGPLAN} Workshop LCTES'98, Montreal, Canada, June 1998, Proceedings}, pages
  16--30, 1998.

\bibitem{FerdinandW99}
Christian Ferdinand and Reinhard Wilhelm.
\newblock Efficient and precise cache behavior prediction for real-time
  systems.
\newblock {\em Real-Time Systems}, 17(2-3):131--181, 1999.

\bibitem{goldreich1996}
Oded Goldreich and Rafail Ostrovsky.
\newblock Software protection and simulation on oblivious rams.
\newblock {\em Journal of the ACM}, 43(3):431--473, 1996.

\bibitem{GrabherGP07}
Philipp Grabher, Johann Gro{\ss}sch{\"a}dl, and Dan Page.
\newblock Cryptographic side-channels from low-power cache memory.
\newblock In {\em International Conference on Cryptography and Coding}, pages
  170--184, 2007.

\bibitem{Gullasch2011}
David Gullasch, Endre Bangerter, and Stephan Krenn.
\newblock Cache games--bringing access-based cache attacks on aes to practice.
\newblock In {\em IEEE Symposium on Security and Privacy}, pages 490--505,
  2011.

\bibitem{GuoKW16}
Shengjian Guo, Markus Kusano, and Chao Wang.
\newblock {Conc-iSE}: Incremental symbolic execution of concurrent software.
\newblock In {\em IEEE/ACM International Conference On Automated Software
  Engineering}, 2016.

\bibitem{GuoKWYG15}
Shengjian Guo, Markus Kusano, Chao Wang, Zijiang Yang, and Aarti Gupta.
\newblock Assertion guided symbolic execution of multithreaded programs.
\newblock In {\em ACM SIGSOFT Symposium on Foundations of Software
  Engineering}, pages 854--865, 2015.

\bibitem{GuoWW17}
Shengjian Guo, Meng Wu, and Chao Wang.
\newblock Symbolic execution of programmable logic controller code.
\newblock In {\em ACM SIGSOFT Symposium on Foundations of Software
  Engineering}, pages 326--336, 2017.

\bibitem{GuoWW18}
Shengjian Guo, Meng Wu, and Chao Wang.
\newblock Adversarial symbolic execution for detecting concurrency-related
  cache timing leaks.
\newblock 2018.

\bibitem{HedinS05}
Daniel Hedin and David Sands.
\newblock Timing aware information flow security for a javacard-like bytecode.
\newblock {\em Electr. Notes Theor. Comput. Sci.}, 141(1):163--182, 2005.

\bibitem{Hu91}
Wei{-}Ming Hu.
\newblock Reducing timing channels with fuzzy time.
\newblock In {\em {IEEE} Symposium on Security and Privacy}, pages 8--20, 1991.

\bibitem{JiangFK16}
Zhen~Hang Jiang, Yunsi Fei, and David~R. Kaeli.
\newblock A complete key recovery timing attack on a {GPU}.
\newblock In {\em IEEE International Symposium on High Performance Computer
  Architecture}, pages 394--405, 2016.

\bibitem{Kocher2018spectre}
Paul Kocher, Daniel Genkin, Daniel Gruss, Werner Haas, Mike Hamburg, Moritz
  Lipp, Stefan Mangard, Thomas Prescher, Michael Schwarz, and Yuval Yarom.
\newblock Spectre attacks: Exploiting speculative execution.
\newblock {\em ArXiv e-prints}, January 2018.

\bibitem{Kocher96}
Paul~C Kocher.
\newblock Timing attacks on implementations of {Diffie-Hellman, RSA, DSS, and
  other systems}.
\newblock In {\em Annual International Cryptology Conference}, pages 104--113.
  Springer, 1996.

\bibitem{KocherJJ99}
Paul~C. Kocher, Joshua Jaffe, and Benjamin Jun.
\newblock Differential power analysis.
\newblock In {\em International Cryptology Conference}, pages 388--397, 1999.

\bibitem{KopfD09}
Boris K{\"{o}}pf and Markus D{\"{u}}rmuth.
\newblock A provably secure and efficient countermeasure against timing
  attacks.
\newblock In {\em {IEEE} Computer Security Foundations Symposium}, pages
  324--335, 2009.

\bibitem{KopfM07}
Boris K{\"{o}}pf and Heiko Mantel.
\newblock Transformational typing and unification for automatically correcting
  insecure programs.
\newblock {\em Int. J. Inf. Sec.}, 6(2-3):107--131, 2007.

\bibitem{KopfMO12}
Boris K{\"o}pf, Laurent Mauborgne, and Mart{\'\i}n Ochoa.
\newblock Automatic quantification of cache side-channels.
\newblock In {\em International Conference on Computer Aided Verification},
  pages 564--580, 2012.

\bibitem{KopfS10}
Boris K{\"{o}}pf and Geoffrey Smith.
\newblock Vulnerability bounds and leakage resilience of blinded cryptography
  under timing attacks.
\newblock In {\em {IEEE} Computer Security Foundations Symposium}, pages
  44--56, 2010.

\bibitem{Lipp2018meltdown}
Moritz Lipp, Michael Schwarz, Daniel Gruss, Thomas Prescher, Werner Haas,
  Stefan Mangard, Paul Kocher, Daniel Genkin, Yuval Yarom, and Mike Hamburg.
\newblock Meltdown.
\newblock {\em ArXiv e-prints}, January 2018.

\bibitem{liu2015}
Chang Liu, Austin Harris, Martin Maas, Michael Hicks, Mohit Tiwari, and Elaine
  Shi.
\newblock Ghostrider: A hardware-software system for memory trace oblivious
  computation.
\newblock {\em ACM SIGARCH Computer Architecture News}, 43(1):87--101, 2015.

\bibitem{liu2016}
Fangfei Liu, Qian Ge, Yuval Yarom, Frank Mckeen, Carlos Rozas, Gernot Heiser,
  and Ruby~B Lee.
\newblock Catalyst: Defeating last-level cache side channel attacks in cloud
  computing.
\newblock In {\em IEEE International Symposium On High Performance Computer
  Architecture}, pages 406--418, 2016.

\bibitem{MangardOP07}
Stefan Mangard, Elisabeth Oswald, and Thomas Popp.
\newblock {\em Power Analysis Attacks - Revealing the Secrets of Smart Cards}.
\newblock Springer, 2007.

\bibitem{MantelS15}
Heiko Mantel and Artem Starostin.
\newblock Transforming out timing leaks, more or less.
\newblock In {\em European Symposium on Research in Computer Security}, pages
  447--467, 2015.

\bibitem{Millen87}
Jonathan~K. Millen.
\newblock Covert channel capacity.
\newblock In {\em {IEEE} Symposium on Security and Privacy}, pages 60--66,
  1987.

\bibitem{MolnarPSW05}
David Molnar, Matt Piotrowski, David Schultz, and David Wagner.
\newblock The program counter security model: Automatic detection and removal
  of control-flow side channel attacks.
\newblock In {\em International Conference on Information Security and
  Cryptology}, pages 156--168. Springer, 2005.

\bibitem{MossOPT12}
Andrew Moss, Elisabeth Oswald, Dan Page, and Michael Tunstall.
\newblock Compiler assisted masking.
\newblock In {\em International Conference on Cryptographic Hardware and
  Embedded Systems}, pages 58--75, 2012.

\bibitem{Mowery2012}
Keaton Mowery, Sriram Keelveedhi, and Hovav Shacham.
\newblock Are aes x86 cache timing attacks still feasible?
\newblock In {\em ACM Workshop on Cloud computing security}, pages 19--24,
  2012.

\bibitem{nagami2008}
Yoshitaka Nagami, Daisuke Miyamoto, Hiroaki Hazeyama, and Youki Kadobayashi.
\newblock An independent evaluation of web timing attack and its
  countermeasure.
\newblock In {\em International Conference on Availability, Reliability and
  Security}, pages 1319--1324, 2008.

\bibitem{OsvikST06}
Dag~Arne Osvik, Adi Shamir, and Eran Tromer.
\newblock Cache attacks and countermeasures: The case of {AES}.
\newblock In {\em Topics in Cryptology - {CT-RSA} 2006, The Cryptographers'
  Track at the {RSA} Conference 2006, San Jose, CA, USA, February 13-17, 2006,
  Proceedings}, pages 1--20, 2006.

\bibitem{page2005}
Dan Page.
\newblock Partitioned cache architecture as a side-channel defence mechanism.

\bibitem{PearceKH04}
David~J. Pearce, Paul H.~J. Kelly, and Chris Hankin.
\newblock Efficient field-sensitive pointer analysis for {C}.
\newblock In {\em {ACM} {SIGPLAN-SIGSOFT} Workshop on Program Analysis For
  Software Tools and Engineering}, 2004.

\bibitem{PhanBPMB17}
Quoc{-}Sang Phan, Lucas Bang, Corina~S. Pasareanu, Pasquale Malacaria, and
  Tevfik Bultan.
\newblock Synthesis of adaptive side-channel attacks.
\newblock In {\em {IEEE} Computer Security Foundations Symposium}, pages
  328--342, 2017.

\bibitem{PierroHW11}
Alessandra~Di Pierro, Chris Hankin, and Herbert Wiklicky.
\newblock Probabilistic timing covert channels: to close or not to close?
\newblock {\em Int. J. Inf. Sec.}, 10(2):83--106, 2011.

\bibitem{rane2015}
Ashay Rane, Calvin Lin, and Mohit Tiwari.
\newblock Raccoon: closing digital side-channels through obfuscated execution.
\newblock In {\em USENIX Security Symposium}, pages 431--446, 2015.

\bibitem{schinzel2011}
Sebastian Schinzel.
\newblock An efficient mitigation method for timing side channels on the web.
\newblock In {\em International Workshop on Constructive Side-Channel Analysis
  and Secure Design}, 2011.

\bibitem{AppliedCryp}
Bruce Schneier.
\newblock {\em Applied cryptography: protocols, algorithms, and source code in
  C}.
\newblock John Wiley \& Sons, 2007.

\bibitem{Shannon48}
Claude~E. Shannon.
\newblock A mathematical theory of communication.
\newblock {\em The Bell System Technical Journal}, 27:379--423, 1948.

\bibitem{Smith09}
Geoffrey Smith.
\newblock On the foundations of quantitative information flow.
\newblock In {\em International Conference on the Foundations of Software
  Science and Computational Structures}, pages 288--302, 2009.

\bibitem{SousaD16}
Marcelo Sousa and Isil Dillig.
\newblock Cartesian hoare logic for verifying k-safety properties.
\newblock In {\em ACM SIGPLAN Conference on Programming Language Design and
  Implementation}, pages 57--69, 2016.

\bibitem{Spreitzer2013}
Raphael Spreitzer and Thomas Plos.
\newblock On the applicability of time-driven cache attacks on mobile devices.
\newblock In {\em International Conference on Network and System Security},
  pages 656--662. Springer, 2013.

\bibitem{stefanov2013}
Emil Stefanov, Marten Van~Dijk, Elaine Shi, Christopher Fletcher, Ling Ren,
  Xiangyao Yu, and Srinivas Devadas.
\newblock Path {ORAM}: an extremely simple oblivious {RAM} protocol.
\newblock In {\em ACM SIGSAC Conference on Computer \& Communications
  Security}, pages 299--310, 2013.

\bibitem{TouzeauMMR17}
Valentin Touzeau, Claire Ma{\"{\i}}za, David Monniaux, and Jan Reineke.
\newblock Ascertaining uncertainty for efficient exact cache analysis.
\newblock In {\em International Conference on Computer Aided Verification},
  pages 22--40, 2017.

\bibitem{vattikonda2011}
Bhanu~C Vattikonda, Sambit Das, and Hovav Shacham.
\newblock Eliminating fine grained timers in xen.
\newblock In {\em ACM workshop on Cloud computing security}, pages 41--46,
  2011.

\bibitem{WangS17}
Chao Wang and Patrick Schaumont.
\newblock Security by compilation: an automated approach to comprehensive
  side-channel resistance.
\newblock {\em ACM {SIGLOG} News}, 4(2):76--89, 2017.

\bibitem{WangWLZW17}
Shuai Wang, Pei Wang, Xiao Liu, Danfeng Zhang, and Dinghao Wu.
\newblock {CacheD}: Identifying cache-based timing channels in production
  software.
\newblock In {\em {USENIX} Security Symposium}, pages 235--252. {USENIX}
  Association, 2017.

\bibitem{wang2007}
Zhenghong Wang and Ruby~B. Lee.
\newblock New cache designs for thwarting software cache-based side channel
  attacks.
\newblock In {\em International Symposium on Computer Architecture}, pages
  494--505, 2007.

\bibitem{YaromF14}
Yuval Yarom and Katrina Falkner.
\newblock {FLUSH+RELOAD:} {A} high resolution, low noise, {L3} cache
  side-channel attack.
\newblock In {\em {USENIX} Security Symposium}, pages 719--732, 2014.

\bibitem{ZhangAM12}
Danfeng Zhang, Aslan Askarov, and Andrew~C Myers.
\newblock Language-based control and mitigation of timing channels.
\newblock In {\em ACM SIGACT-SIGPLAN Symposium on Principles of Programming
  Languages}, pages 99--110, 2012.

\bibitem{ZhangGSW18}
Jun Zhang, Pengfei Gao, Fu~Song, and Chao Wang.
\newblock {SCInfer}: Refinement-based verification of software countermeasures
  against side-channel attacks.
\newblock In {\em International Conference on Computer Aided Verification},
  2018.

\end{thebibliography}

\end{document}